\documentclass[a4paper,fleqn,usenatbib]{mnras}

\usepackage{newtxtext,newtxmath}

\usepackage[T1]{fontenc}
\usepackage{ae,aecompl}

\usepackage{graphicx}	
\usepackage{amsmath}	
\usepackage{amssymb}	

\title[R modes in accreting white dwarfs]{R-mode oscillations in accreting white dwarfs in cataclysmic variables}

\author[H. Saio]{
Hideyuki Saio \thanks{E-mail: saio@astr.tohoku.ac.jp}
\\
Astronomical Institute, Graduate School of Science, Tohoku University, Sendai 980-8578, Japan\\
}

\date{Accepted XXX. Received YYY; in original form ZZZ}

\pubyear{2019}

\begin{document}
\label{firstpage}
\pagerange{\pageref{firstpage}--\pageref{lastpage}}
\maketitle

\begin{abstract}
Some cataclysmic variables show short-period ($\sim$200 to $\sim$2000 s) light variations, which are attributable to nonradial pulsations of the accreting primary white dwarf. 
We regard these periodic variations as r-mode (global Rossby-wave) oscillations which are presumably excited mechanically and confined into hydrogen-rich layers in the white dwarf.
Often observed modulations in amplitudes and frequencies even during an observing season may be interpreted as beatings among densely distributed r-mode frequencies. 
Making the r-mode frequency distribution of a model to be consistent (by adjusting the rotation frequency) with the observed pulsation frequencies, we obtain the best-fit rotation rate for each case. 
The rotation periods thus obtained for 17 accreting white dwarfs lie between 4 and 10 minutes.
Pulsation frequencies observed sometime after a dwarf-nova outburst tend to be higher than the pre-outburst values, so that the best-fit rotation rate for the post-outburst frequencies is also higher. This indicates a spin-up of the hydrogen-rich layers to occur by an enhanced accretion at the outburst.
We also discuss the relation between the rotation frequencies of pulsating  accreting white dwarfs and the orbital periods.
\end{abstract}

\begin{keywords}
binaries:close -- stars:dwarf novae -- stars:low-mass -- stars:oscillations -- stars:rotation -- white dwarfs
\end{keywords}



\section{Introduction} \label{sec:intro}
Short-period light variations of an accreting white dwarf (WD), which are attributable to nonradial pulsations, were first discovered by \citet{war98} in the cataclysmic variable (dwarf nova) GW~Lib at quiescence.
Since then, the number of observed accreting WD pulsators has increased to nearly twenty \citep[see e.g.][for reviews]{szk13pro,szk15,muk17}.
They belong to WZ Sge type cataclysmic variables having very low mass-transfer rates of $\sim\!\!10^{-11}M_\odot$yr$^{-1}$ at quiescence, and have very large outbursts in very long intervals (a few decades).
Because of the very low mass-transfer rate, a large fraction of the flux from a system at quiescence is originated from the WD, so that pulsations of the WD can be detected.

The pulsations of accreting WDs have been generally regarded as g modes excited thermally in the hydrogen and helium ionization zones  
similarly to the g modes 
in the ZZ Ceti variables  \citep[e.g.][]{bri91,gol99,sai13}.
Although the effective temperature tends to be higher than the blue edge of the ZZ Ceti instability strip \citep{szk10}, it is thought that an enhanced helium abundance would solve the problem \citep{arr06,vang15}.

\begin{figure}
\includegraphics[width=\columnwidth]{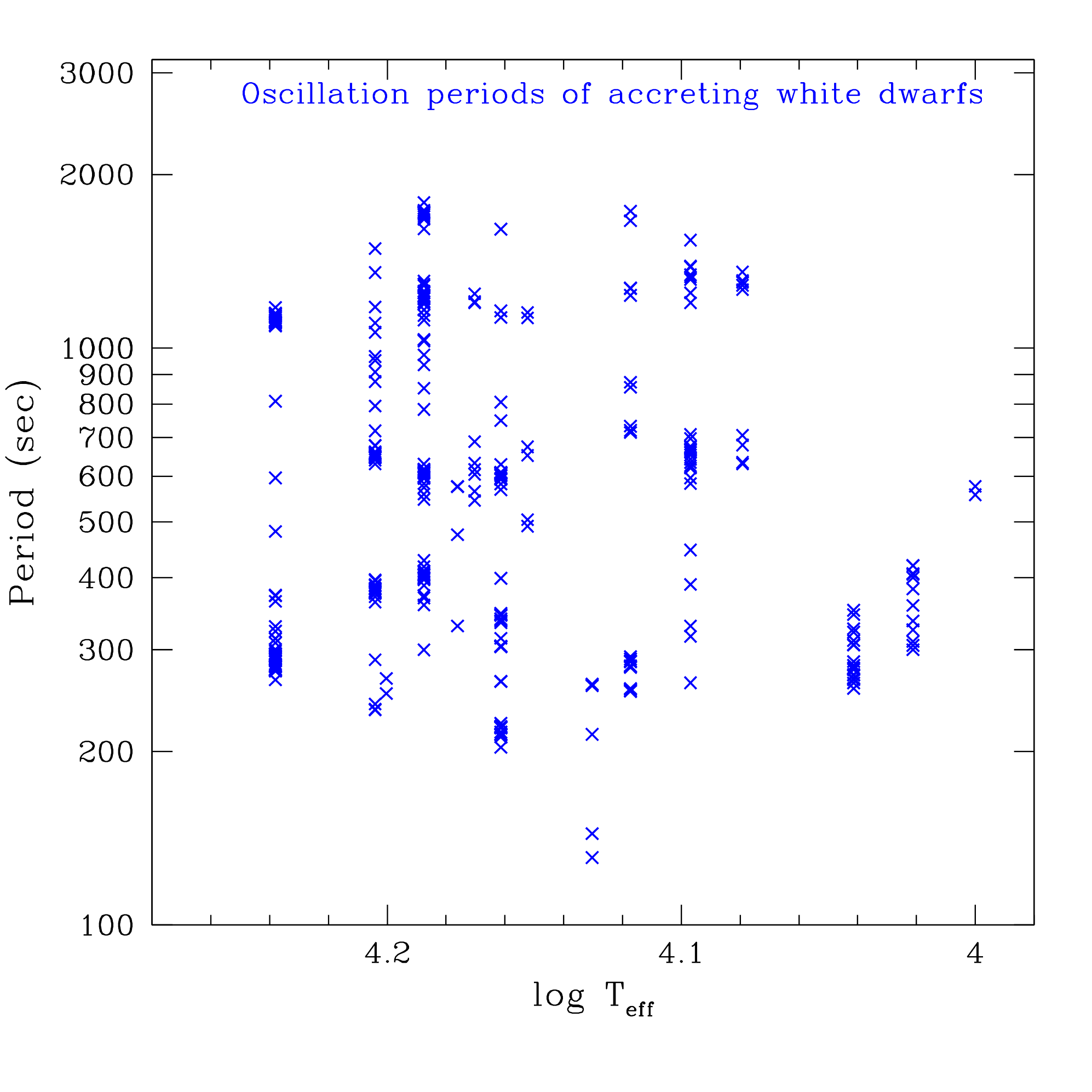}  
\caption{Short periodicities in accreting white dwarfs versus effective temperature. References for the periods and the effective temperature for each star are given in \S\ref{sec:cv} and Appendix. Longer periods ($> 2000$~s) are not shown because they are likely originated from accretion disks or are related with orbital motions. }
\label{fig:periods}
\end{figure}

Fig.~\ref{fig:periods} shows oscillation periods of accreting WDs, which we discuss in the following sections, with respect to the effective temperatures. 
Although the period range is similar to that of ZZ Ceti variables, there is no trend in the periods of accreting WDs with respect to the effective temperature. 
This is a stark contrast to the well-established trend of ZZ Ceti variables that cooler ZZ Ceti stars have longer periods \citep[e.g.][]{muk06,her17}. This suggests   
considerable differences in the oscillation properties between accreting WDs and ZZ Ceti stars.

In addition, there is a significant difference in rotation rates between accreting WDs in cataclysmic variables and ZZ~Ceti variables. 
Rotation periods of ZZ~Ceti variables  (longer than 5~hr in most cases) are much longer than the g-mode periods (100--1500 s) \citep{fon08,kaw15,her17}, so that rotation is only a small perturbation to g-mode oscillations in ZZ~Ceti stars.
 In contrast, accreting WDs rotate much faster,  
having projected rotation velocities larger than 100~km~s$^{-1}$ for most cases \citep[see e.g.][]{sio12}.
Among pulsating accreting WDs, \citet{szk12} obtained a rotation period of 209~s for GW~Lib from the spectroscopic projected rotation velocity $V\sin i=40$~km~s$^{-1}$ combined with an inclination of $11^\circ$ \citep{vanspa10apj}.
For EZ Lyn \citet{szk13ez} obtained $V\sin i=225\pm75$~km~s$^{-1}$, which corresponds to a rotation period less than 350~s for a typical WD radius. 
These rotation periods are comparable to or shorter than the oscillation periods of these stars. 
 Therefore, the effects of the Coriolis force should be significant, and
g-mode characters should be very different from those in ZZ~Ceti variables.

It is known that, among g modes, prograde sectoral modes are predominantly observed in rapidly rotating main-sequence g-mode pulsators \citep[e.g.,][]{vanr16,oua17}.
This is explained by the fact that if the pulsation frequency of a star in the co-rotating frame is less than the rotation frequency, a g mode other than the prograde sectoral modes would have amplitude strongly confined into a narrow equatorial zone \citep[e.g.,][]{sai18g,sai18pro}, and hence should have low visibility due to geometrical cancellation.
Therefore, if pulsations in an accreting WD were g modes, they should be prograde sectoral modes.
In the inertial (observer's) frame, however, a prograde mode should have a period shorter than the rotation period, which contradicts to the fact that most periods observed in GW Lib  are {\it longer} than the rotation period \citep{szk12}. 

Thus, the observed pulsations must be retrograde (in the co-rotating frame) modes.
However, as noted above, retrograde g modes should be hardly detected due to the cancellation on the surface.  
In rapidly rotating $\gamma$ Dor stars, pulsations whose periods longer than the rotation period are identified as r modes (global Rossby waves) \citep{vanr16}. (The dependence of period-spacing on period clearly distinguishes between g and r modes.)
Therefore, it seems more sensible to consider r modes in accreting WDs rather than g modes. 
   
In this paper we consider the pulsations detected in accreting WDs in cataclysmic variables to be r-mode oscillations.
The next section (\S\ref{sec:rmodes}) summarises the property of r-mode oscillations in WDs. 
In \S\ref{sec:cv} (and Appendix) we fit observed pulsation frequencies for each star (in each epoch in some cases) with the frequency distribution of r modes calculated for a WD model with an assumed rotation frequency. In other words, we determine the best-fit rotation frequency for each case.  \S\ref{sec:discussion} discusses the relation between rotation frequencies of accreting WDs and the orbital periods of the cataclysmic binaries.

\section{R-mode oscillations in white dwarfs}
\label{sec:rmodes}
The first clear signature of r-mode oscillations was found by \citet{vanr16} in some rapidly rotating $\gamma$ Dor variables as the period-spacing--period relation having a gradient opposite to that of g modes. 
Later, from the properties of visibility-frequency relations of r modes, \citet{sai18} identified r modes in various main-sequence stars; r-mode oscillations seem to be commonly present in non-supergiant stars. 
General properties of r modes are discussed in \citet{sai18,sai18pro}.
Here we discuss only the properties to be used in the following section. 

In studying low-frequency oscillations of a rotating star,  the traditional approximation of rotation (TAR) is useful, in which the horizontal component of the angular velocity of rotation $\Omega\sin\theta$ ($\theta$ is co-latitude) and the Eulerian perturbation of gravitational potential are neglected.
Then, the governing equations are reduced to the same equations for nonradial pulsations of a non-rotating star under the Cowling approximation, except that $\ell(\ell+1)$ is replaced with $\lambda$, the  eigenvalue of the Laplace Tidal equation, and the latitudinal dependence is given by the eigenfunction $\Theta^m_k(\cos\theta)$ instead of the Legendre function $P_\ell^m(\cos\theta)$ \citep[see e.g.][]{lee97}.
Both $\lambda$ and the functional form of $\Theta^m_k$ vary depending on the spin parameter, $s$,  defined as
\begin{equation}
s \equiv {2\Omega\over(2\upi\nu^{\rm co})} = {2f_{\rm rot}\over \nu^{\rm co}},
\end{equation} 
where $\nu^{\rm co}$ is the cyclic frequency of pulsation in the co-rotating frame, and $f_{\rm rot}$ is the cyclic frequency of rotation.
The impact of Coriolis forces on pulsation is significant if $s > 1$.

Because of the similarity of the differential equations, we can formally use an asymptotic expression for g-mode frequencies in a non-rotating star for low-frequency g and r modes in a rotating star by replacing $\ell(\ell+1)$ with $\lambda$ as
\begin{equation} 
\nu^{\rm co} \approx {\sqrt{\lambda}\over 2\upi^2n_{\rm g}}\int{N\over r}dr \equiv {\sqrt{\lambda}\over n_{\rm g}}\nu_0,
\label{eq:nuco}
\end{equation}
where $N$ is the Brunt-V\"ais\"al\"a frequency \citep[e.g.][]{unno,ack10}, $n_{\rm g}$ the number of radial nodes, and $\int dr$ means integration through the propagation layers.

Although the equation is formally similar to the non-rotating case, variation of $\lambda$ as a function of the spin parameter significantly  modifies the properties of low-frequency non-radial pulsations if $s>1$ ( in particular for retrograde modes). 
One of the most significant influence of the Coriolis force is the emergence of r modes (global Rossby waves coupled with buoyancy) in a rotating star.
In the retrograde side of a $s$-$\lambda$ diagram \citep[e.g.,][]{lee97,sai18pro}, a group of lines appear in the positive range of $\lambda$ for $s\ge 2$. 
These lines represent $\lambda$ for r modes which are ordered by using negative integer $k$ ($= -1, -2, -3 \ldots$).
These r modes are present (i.e., $\lambda > 0$) if
\begin{equation}
2\pi\nu^{\rm co}(\mbox{r modes}) < {2m\Omega\over(m+|k|)(m+|k|-1)} \le \Omega = 2\upi f_{\rm rot}
\label{eq:ineq}
\end{equation}
with a positive $m$. (We adopt the convention, as in our previous papers, that a positive $m$ corresponds to retrograde modes.)

\begin{figure}
\begin{center}
\includegraphics[width=0.32\columnwidth]{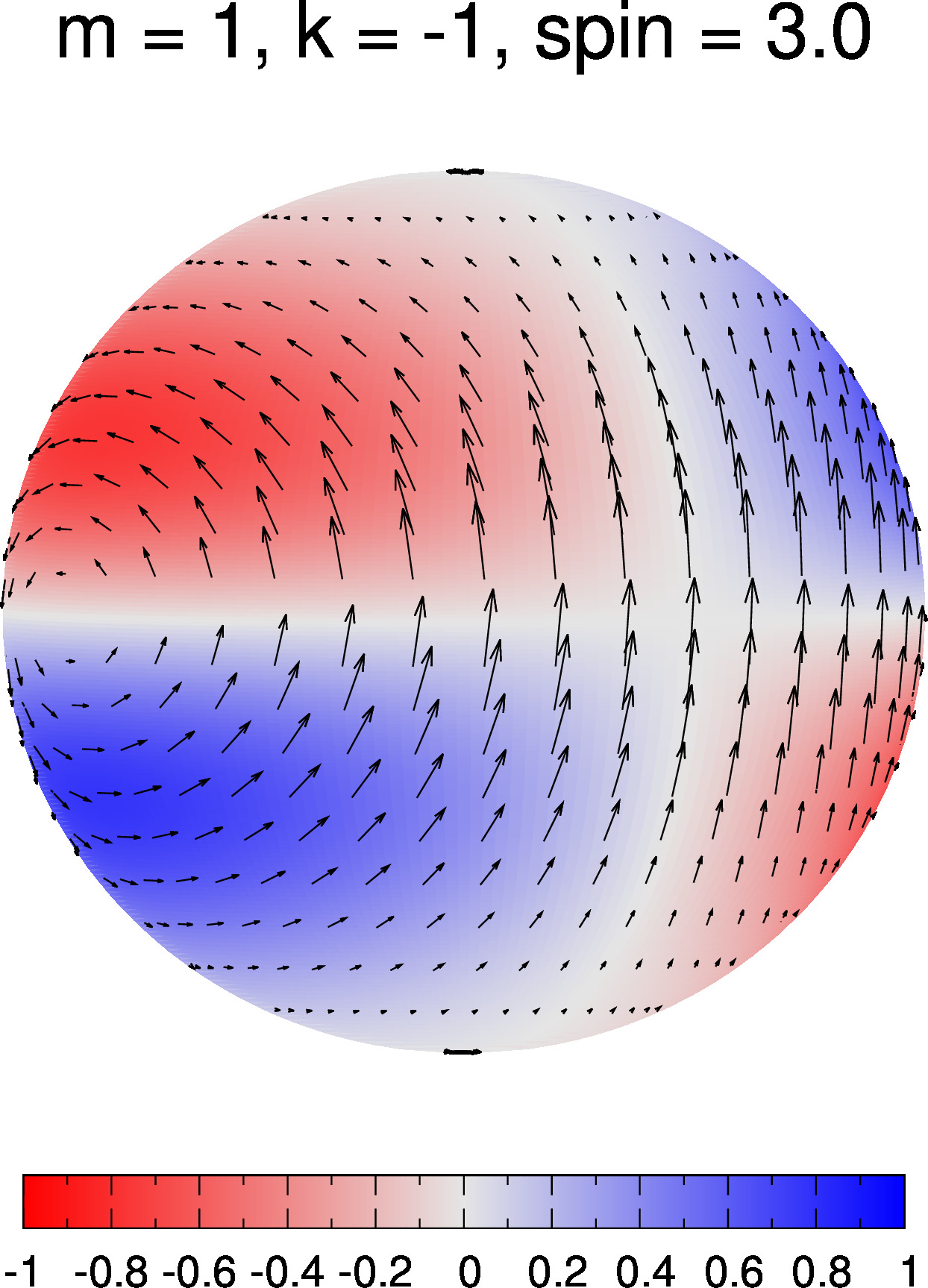} 
\includegraphics[width=0.32\columnwidth]{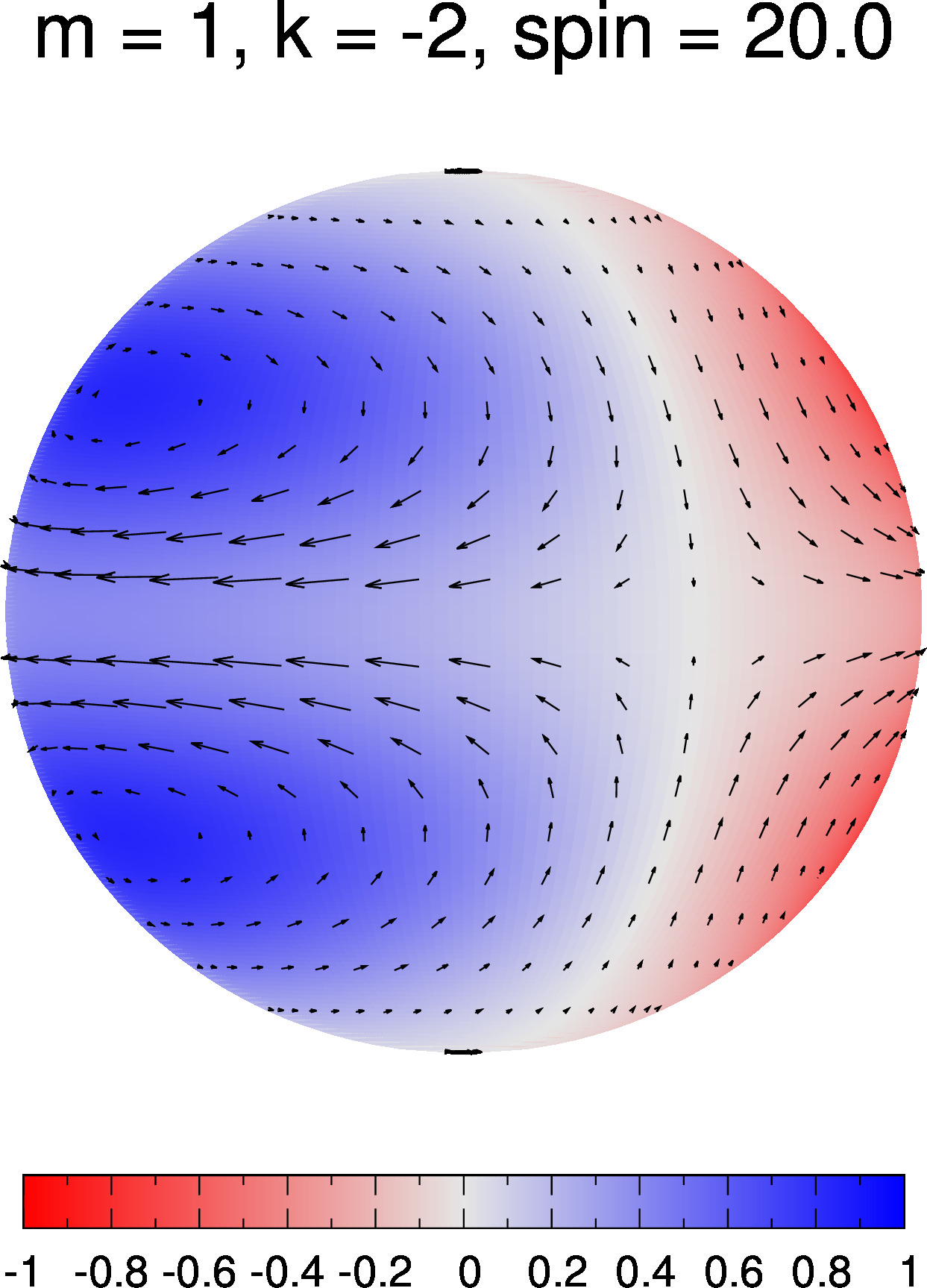} 
\includegraphics[width=0.32\columnwidth]{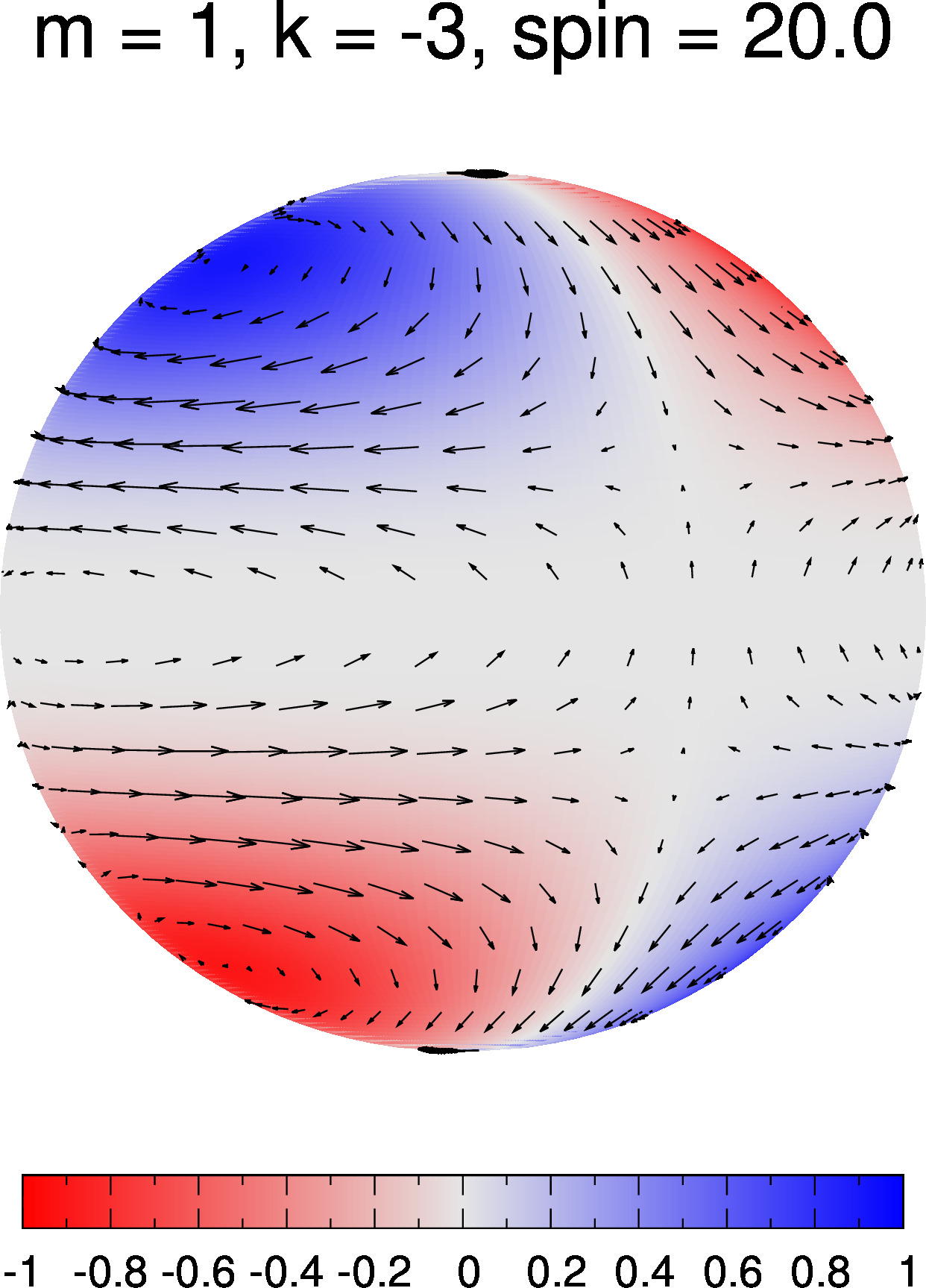} 
\end{center}
\caption{Flow vectors and temperature perturbations on the stellar surface for r modes of $m=1$ with $k=-1,-2,$ and $-3$ at indicated spin parameters.}
\label{fig:pattern}
\end{figure}

While $\lambda$ for $k=-1$ increases rapidly as the spin parameter, $s$, increases, $\lambda$ for $k\le -2$ stay low as represented by the asymptotic form
\begin{equation}
\lambda \approx {m^2\over(2|k|-1)^2} \quad {\rm if} \quad k\le -2 ~ \& ~ s\gg 1 
\label{eq:lambda}
\end{equation}
derived by \citet{tow03a}.
Fig.~\ref{fig:pattern} shows snapshots of temperature and velocity distributions of $m=1$ r modes with $k=-1,-2,$ and $-3$ on the surface for selected spin parameters.
The number of vortices in a (east or west) hemisphere is given by $|k|$, while an even (odd) $|k|$ corresponds to temperature perturbations symmetric (antisymmetric) to the equator.  

\begin{figure}
\includegraphics[width=\columnwidth]{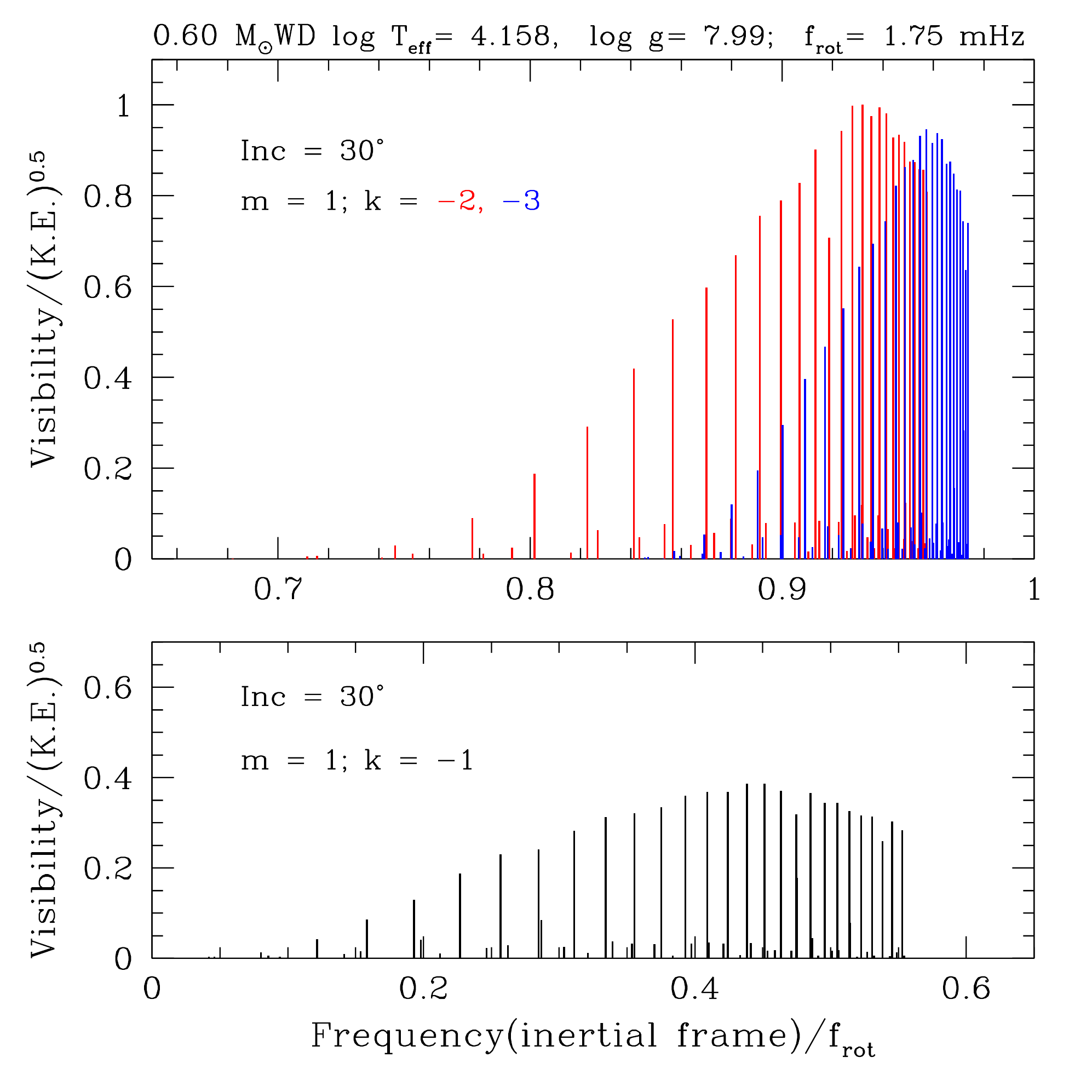}  
\caption{Visibility of each r mode of $m=1$ versus its frequency in the inertial frame normalized by the rotation frequency. Temperature variations of r modes with $k=-2$ are symmetric (even) with respect to the equator, while antisymmetric (odd) with $k=-1$ and $-3$. The visibility is normalized as unity at the maximum. The visibility of a mode non-trapped to the hydrogen-rich layers is very small, while trapped modes tend to have large visibilities.}
\label{fig:vis}
\end{figure}

Since $\nu^{\rm co} < f_{\rm rot}$  (eq.\ref{eq:ineq}), r-mode frequencies in the inertial frame, $\nu^{\rm inert}$, are given as
\begin{equation}
\nu^{\rm inert} = mf_{\rm rot}-\nu^{\rm co} \approx mf_{\rm rot}-{\sqrt{\lambda}\nu_0\over n_{\rm g}} \qquad \mbox{for r modes}.
\end{equation}
Thus, frequencies of r modes with azimuthal order $m$ should be observed between $(m-1)f_{\rm rot}$ and $mf_{\rm rot}$. 
Fig.~\ref{fig:vis} shows frequency ranges of $m=1$ r modes for $k=-1,-2,$ and $-3$ with visibilities under the assumption of energy equi-partition \citep[see][for details]{sai18} for an inclination of $30^\circ$.
Generally, $k=-2$ modes (symmetric) have larger visibilities, while visibilities of odd modes ($k=-1,-3$) decrease with increasing the inclination.

\begin{figure}
\includegraphics[width=\columnwidth]{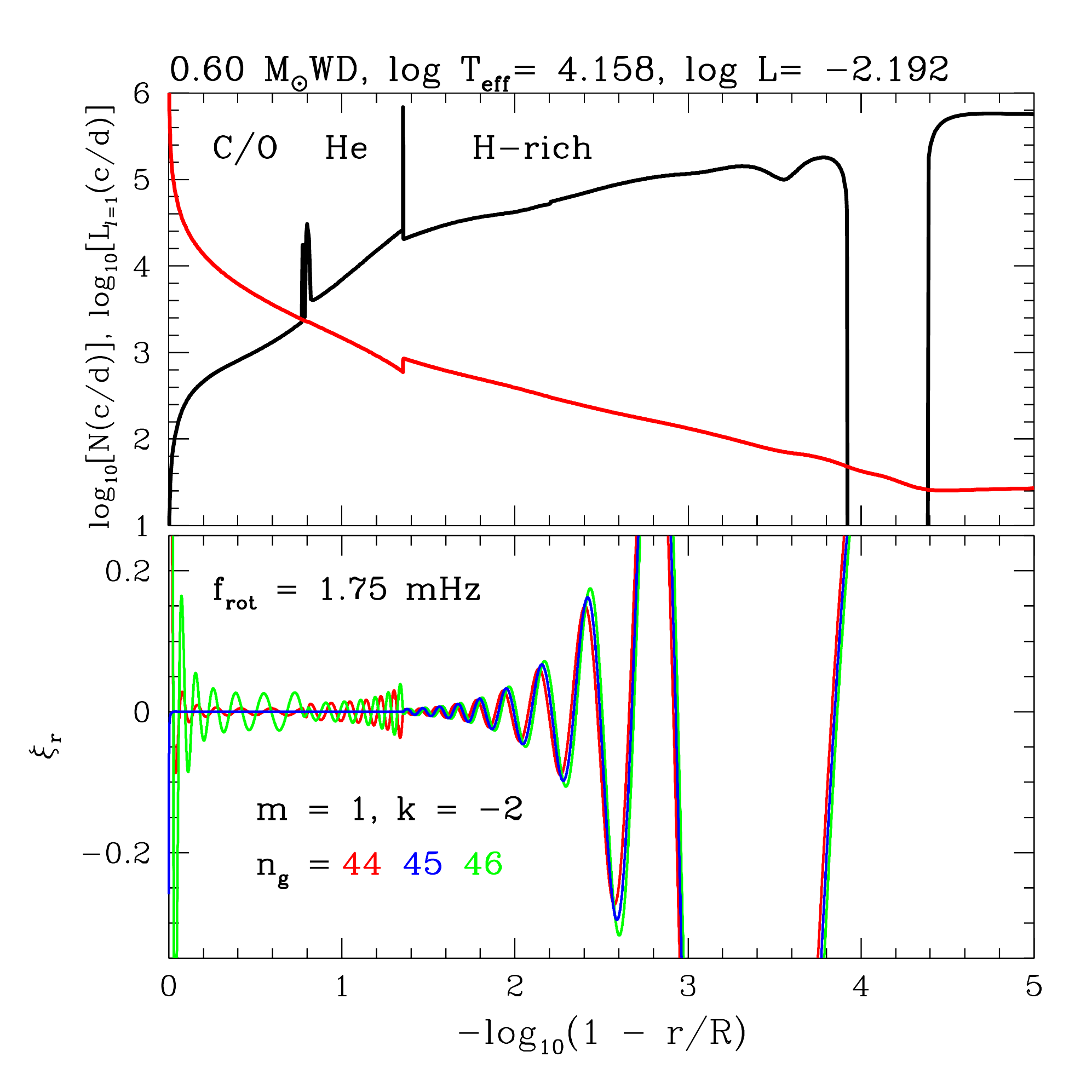}  
\caption{{\bf Upper panel:} A propagation diagram in a non-rotating WD with a mass of $0.6~M_\odot$. Black line is for the Brunt-V\"ais\"al\"a frequency, $N$, while red line for the Lamb frequency $L_{\ell=1}$ \citep[see e.g.,][for the definitions]{unno}.
Sharp peaks in the Blunt-V\"ais\"al\"a frequency occur by steep gradients of chemical composition at the interfaces between H-rich and He layers and between the He layers and the C/O degenerate core. 
{\bf Lower panel:} Runs of the radial displacement $\xi_r$ for some even ($k=-2$) r modes of $m=1$. 
R-mode oscillations can be reflected at the bottom of the H-rich layers and trapped in the outer layers as the mode with $n_g=45$ (blue line), while the modes of $n_g= 44, 46$ penetrate into the center because
reflections at the H:He and the He:C/O interfaces are weak.
}
\label{fig:trapping}
\end{figure}

As the number of radial nodes ($n_{\rm g}$) increases, $\nu^{\rm co}$ decreases (eq.\ref{eq:nuco}) so that $\nu^{\rm inert}$ increases approaching $mf_{\rm rot}$, and frequency spacing becomes smaller (see Fig.~\ref{fig:vis}).
Each sequence stops, before reaching $mf_{\rm rot}$, because $\nu^{\rm co}$ must be larger than the critical frequency, $\nu_{\rm crit}$ (effective Lamb frequency at the stellar surface, $r=R$; see upper panel of Fig.~\ref{fig:trapping}); i.e.,
\begin{equation}
\nu^{\rm co} > \nu_{\rm crit} = {\sqrt{\lambda}c_{\rm s}\over 2\upi R} 
\end{equation} 
with $c_{\rm s}$ being sound speed at the surface \citep{tow00}.
Among the cases of $k=-1,-2,-3$ the frequency gap to $mf_{\rm rot}$ (i.e., $\nu_{\rm crit}$) is smallest for $k=-3$ because the limiting value of $\lambda$ is smallest (eq.~\ref{eq:lambda}), while the gap is larger for hotter stars because of higher sound speed at the surface.

Fig.~\ref{fig:vis} also shows that  visibilities of some modes are very small compared to other ones at similar frequencies.
The difference is caused by whether the mode is trapped in the H-rich layers or penetrates into deeper layers.
The lower panel of Fig.~\ref{fig:trapping} shows radial displacement $\xi_r$ for a trapped mode ($n_{\rm g} = 45$; blue line) and two non-trapped modes ($n_{\rm g}=44, 46$). 
Since matter density is very high in deeper layers, non-trapped modes should have higher kinetic energy compared with modes trapped in the H-rich layers for a given surface amplitude. (For all cases $\xi_r$ is normalized as unity at the surface.)
So, if energy is equally given to each mode, trapped modes should have larger amplitude at the surface and hence have larger visibilities.
For the cases shown in the lower panel of Fig.~\ref{fig:trapping}, the visibility of the trapped mode ($n_{\rm g} = 45$) is more than 20 times larger than the two non-trapped modes.
Thus, only r modes trapped in the H-rich layers should be observable.

\section{R-mode fittings to observed pulsations}
\label{sec:cv}
While the possibility of r modes was briefly discussed in \citet{muk13}, pulsations in accreting white dwarfs were generally thought as g modes similar to those in ZZ Ceti variables in the literature.
As noted in \S~\ref{sec:intro}, however, g modes in rapidly rotating stars have characters inconsistent with observed pulsations in accreting (and hence rapidly rotating) white dwarfs. 
 
In this section, we fit the theoretical frequency range of r mode oscillations  to pulsation frequencies observed in various accreting WDs in cataclysmic variables.
The most important parameter for r-mode frequencies is the rotation frequency ($f_{\rm rot}$), while details of the interior structure are not very important. 
To calculate the r-mode frequencies and visibilities for a rotation frequency, we have used a static white dwarf model ($0.6$ or $0.8~M_\odot$) having an effective temperature roughly consistent with a spectroscopically determined value for each case.

Utilized white dwarf models have masses of helium and H-rich (solar abundance) layers $(M_{\rm He}/M_\odot,M_{\rm H}/M_\odot) = (8\times10^{-3},7\times10^{-6})$ and $(2\times10^{-3},3\times10^{-6})$ for the $0.6$ and $0.8~M_\odot$ models, respectively.  
The cooling evolution (without rotation and atomic diffusion) for each mass was  calculated after  putting solar-abundance matter of $M_{\rm H}$ on the He-accreating steady-state model (with $10^{-7}~M_\odot$~yr$^{-1}$) taken from \citet{kat18} (the accretion was stopped at the start of the cooling calculation). 
A propagation diagram of a $0.6~M_\odot$ model is shown as an example in the upper panel of Fig.~\ref{fig:trapping}.

\subsection{GW Lib}
\label{sub:gwlib}
Short-period light variations indicative of nonradial pulsations of the accreting primary WD in a cataclysmic variable (CV) were first detected in GW Lib by \citet{war98}.
Since then, many observations have been performed; 
GW Lib is the most studied case among pulsating accreting WDs. In addition, GW Lib is one of the few cases where pulsations were observed during quiescence before and after an outburst.  
GW Lib had a very large (9~magnitude) outburst in 2007 after 24 years of the discovery in 1983 \citep{tem07}. 
The pulsation signals disappeared at the outburst, but they seemed to return one year after the outburst \citep{cop09} at shifted frequencies.  

\begin{figure}
	\includegraphics[width=\columnwidth]{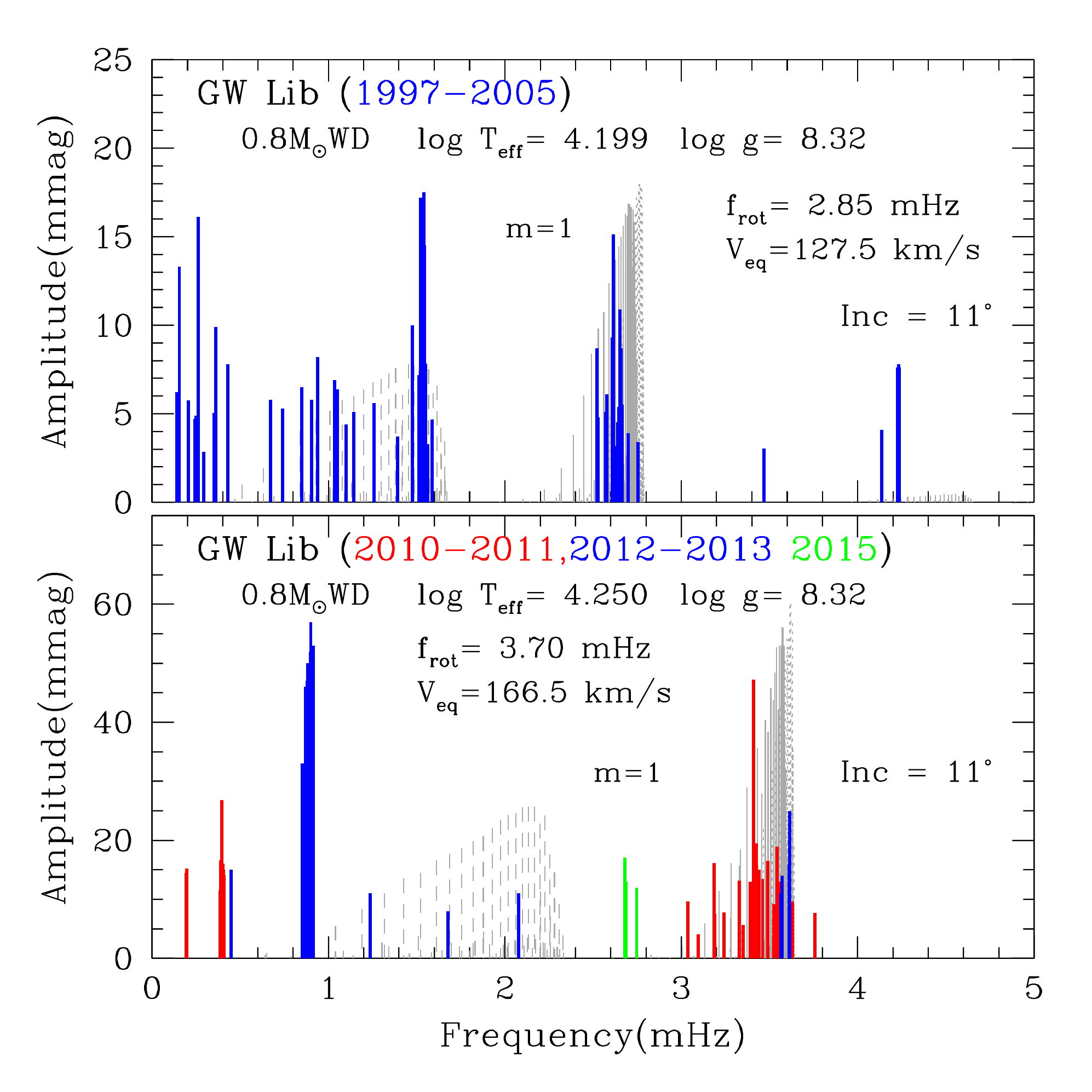}  
    \caption{Frequencies of periodic light variations of GW Lib observed before (upper panel) and after (lower panel) the outburst in 2007.
    Observed year of each frequency is colour coded as indicated.
Gray lines show theoretical visibilities at r-mode frequencies of $m=1$; solid lines are for even modes ($k=-2$), and long and short dashed lines are for odd modes of $k=-1$ and $k=-3$, respectively. 
The maximum visibility in each panel is normalized arbitrary.
 Rotation frequencies ($f_{\rm rot}$) chosen to fit observed main frequency groups with $m=1$ r modes are $2.85$~mHz and $3.70$~mHz before and after the outburst, respectively.}
    \label{fig:gwlib}
\end{figure}

Fig.~\ref{fig:gwlib} compares r-mode frequencies/visibilities with observed frequencies/amplitudes of GW~Lib.
The upper panel shows frequencies prior to the outburst, taking from  \citet{vanz04} (1997--2001), \citet{szk02gw} (2002 January), and \citet{cop09}(2005 May). The lower panel shows pulsation frequencies/amplitudes after the outburst obtained by \citet{szk12}(2010 March -- 2011 August), \citet{cho16}(2012 May), \citet{szk15} (2012 June, 2013 March), and \citet{szk16} (2015 April).
 Each observing run detects only few frequencies which are often similar to but slightly different from the frequencies previously obtained. This is a common property of short periods variations of this type (i.e. accreting WDs).
This can be explained as beating among densely distributed r-mode frequencies (Fig.~\ref{fig:vis}), and a poor frequency resolution due to a short time baseline.
For this reason, all frequencies/amplitudes reported in the literature are included in this figure (and in the following cases).

Gray lines show theoretical visibilities of r-mode oscillations for $0.8$-$M_\odot$ WD models.  
(cf. \citet{vanspa10apj} obtained $0.84~M_\odot$.)
A slightly higher effective temperature is adopted for the model after the outburst, taking into account the accretion heating \citep{szk12}, although the property of r modes is not  very sensitive to $T_{\rm eff}$.

To fit r-mode frequency ranges to observed main frequency groups of GW~Lib, we have chosen rotation frequencies of $2.85$~mHz prior to the 2007 outburst and $3.70$~mHz after the outburst (rotation periods of $351$~s and $270$~s, respectively).
These rotation frequencies should represent rotation rates in the H-rich envelope, because visible r modes are confined there.
(In other words, the rotation rate of the WD core could be considerably different from that in the outer envelope.)
Fig.~\ref{fig:gwlib} shows that main frequency groups of GW~Lib are roughly consistent with $m=1$ r modes, while $\sim\!\!4.2$~mHz frequencies in the upper panel can be explained as combination frequencies between $\sim\!\!1.5$~mHz and $\sim\!\!2.6$~mHz  \citep{vanz04}.
The $3.47$~mHz frequency in the upper panel of Fig.~\ref{fig:gwlib} cannot be explained. The frequency was only marginally detected in 1998 June data \citep{vanz04}. It may be a high-order combination frequency or a transient feature.

Low frequency features at $\sim\!\!0.43-0.2$~mHz should be related with the orbital frequency $0.217$~mHz \citep{tho02} and its harmonic and also with superhump frequencies. 
The strong feature at $\sim\!\!0.9$~mHz (19~min; blue lines in the lower panel) appeared in optical (but not in UV) observations in 2012 May--June  \citep{szk15,cho16}, also in 2008 April--June \citep{cop09,sch10,bul11}.
Although it is located within the frequency range of $m=1 (k=-1)$ r modes,
the 19~min feature is likely originated from the accretion disk because of its transient nature \citep{bul11}.

Another long-period features at 2 and 4~hr (not shown) were detected time to time \citep{wou02,cop09,hil07,szk16,tol16}.
\citet{tol16} found that these light variations (as well as shorter period ones) are caused by considerable temperature variations.
They argued a possible explanation for these long periodicities by retrograde low-order g modes despite the expected low visibility due to an amplitude confinement toward a narrow equatorial zone. 
Long period r-mode pulsations are free from such amplitude confinement.
However, the longest period of r model is about 2~hr in the model after the outburst, while it is about 4~hr in the model prior to the outburst.
Thus, it does not seem possible to identify 4~hr variability after the outburst as an r mode.
The origin of these long periodicities is unclear \citep[see also the discussion in][]{cop09}. 

Our post-outburst model has an equatorial rotation velocity of  $167$~km~s$^{-1}$.
Combining it with the inclination angle $11^\circ$ \citep{vanspa10apj}, we have $V\sin i = 32$~km~s$^{-1}$.
This is roughly consistent with the spectroscopic estimate $V\sin i = 40$~km~s$^{-1}$ obtained in 2010 by \citet{szk12}.
 
Our r-mode models for the nonradial pulsations of GW~Lib indicate that the H-rich envelope of the white dwarf was considerably spun up by the accretion during the 2007 outburst.
Rotation frequencies $2.85$~mHz and $3.70$~mHz before and after the outburst, respectively, correspond to an increase from $0.033\Omega_{\rm K}$ to $0.043\Omega_{\rm K}$ with $\Omega_{\rm K}$ being the Keplerian angular frequency at the surface of the WD.
This indicates that during the outburst an angular momentum of $\sim\!\!0.01\Omega_{\rm K}(2R^2M_{\rm H}/3)$ was accreted, where $M_{\rm H}$ is the mass of H-rich layers.
Letting $M_{\rm acc}$ to be the mass accreted during the outburst, the balance of angular momentum is approximately given as 
\begin{equation}
\Omega_{\rm K}R^2M_{\rm acc} \approx 0.01\Omega_{\rm K}{2R^2M_{\rm H}\over3} \quad \rightarrow \quad
M_{\rm acc} \approx 0.01 {2M_{\rm H}\over3}.
\label{eq:angm}
\end{equation}
\citet{vic11} derived a bolometric luminosity of $1.5\times10^{34}$ erg~s$^{-1}$ during the 26-day outburst of GW Lib; i.e.,  energy of $\approx3\times10^{40}$~erg was released during the outburst. 
The energy (= gravitational energy of accreted matter) corresponds to $M_{\rm acc} \approx 2\times10^{-10}M_\odot$.
Substituting this into equation~(\ref{eq:angm}), we obtain $M_{\rm H}\approx 3\times10^{-8}M_\odot$. Combining this $M_{\rm H}$ with the mean accretion rate $1.3\times 10^{-11}~M_\odot$~yr$^{-1}$ obtained by \citet{vic11}, we can infer that the last classical nova explosion of GW~Lib occurred about $2\times10^3$ years ago. 

Pulsation frequencies detected in 2015 \citep[][green lines in the lower panel of Fig.~\ref{fig:gwlib}]{szk16} at $\sim\!2.7$~mHz are lower than those observed in $2010 - 2013$, and comparable to the frequencies for even r modes prior to  the outburst (upper panel);
i.e., consistent with the rotation frequency prior to the outburst.
This may indicate accreted angular momentum being re-distributed after the outburst, and rotation frequency is nearing the value prior to the outburst.
Further monitoring of pulsations in GW~Lib would be very interesting.

\subsection{EQ Lyn (= SDSS J074531.92+453829.6)}
\label{sub:eqlyn}
\begin{figure}
	\includegraphics[width=\columnwidth]{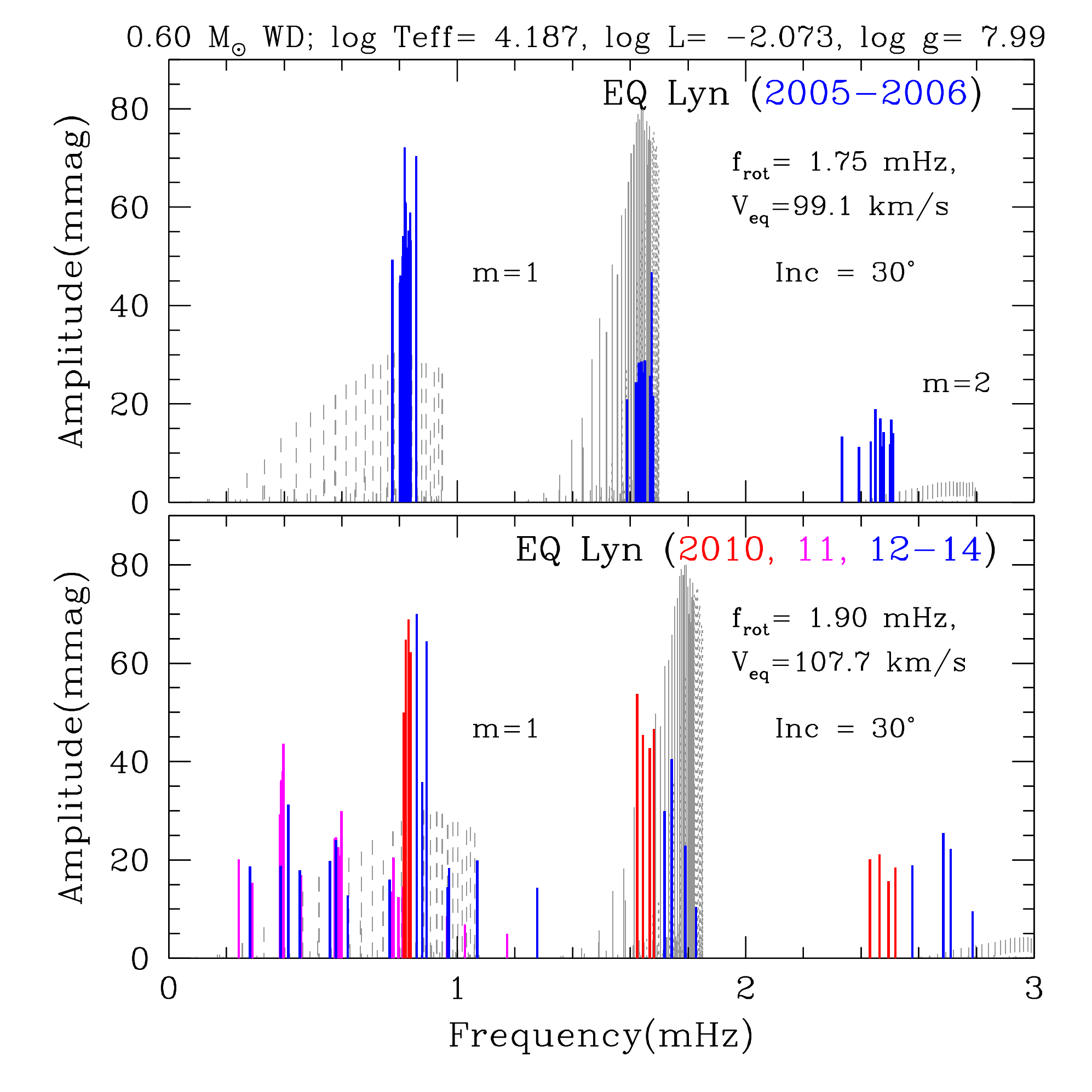}  
    \caption{EQ Lyn before (upper panel) and after (lower panel) the outburst in 2006.
    Frequency/amplitude data before and after the outburst are taken from \citet{muk11,muk13}, and \citet{szk15}.
    Observed years are colour coded as indicated.
    Gray lines indicate visibilities of r modes. 
    In each panel, the visibility is normalized arbitrarily such that the maximum of even $m=1 (k=-2)$ modes corresponds to 0.08 mmag. Since no information on the inclination angle of EQ~Lyn, $30^\circ$ is arbitrary chosen.
    Rotation frequency ($f_{\rm rot}$) for each panel is chosen for the predicted visibility distribution of r modes to be consistent with observed frequencies.
    }
    \label{fig:eqlyn}
\end{figure}
EQ~Lyn was first identified as a CV from the SDSS (Sloan Digital Sky Survey) spectrum by \citet{szk06}. 
\citet{muk07} discovered, in 2005 October, short-period light variations, which are attributed to non-radial pulsations in the primary white dwarf. 
Similar short-period signals were detected 
from 2005 November to 2006 January \citep{muk11}.
EQ Lyn underwent an outburst in 2006 October, which was detected by the Catalina Real-Time Transient Survey (CTRS).
After the outburst no pulsations had been observed for more than three years.
Even after pulsations were detected again in 2010 February--March,  pulsations disappeared time to time; pulsations seem to disappear when superhump signatures appear \citep{muk13}.
The superhump is caused by the precession of a fully developed elliptic accretion disk \citep[e.g.][]{whi88,hir90}.
The cause of the anti-correlation is not clear.

Fig.~\ref{fig:eqlyn} shows observed frequencies and amplitudes detected before (upper panel) and after (lower panel) the outburst in 2006.
Obvious superhump ($0.19$~mHz) and orbital ($0.21$~mHz) frequencies are not shown.
The pulsation frequencies/amplitudes are compared with r-mode visibilities  for a $0.6~M_\odot$ WD model with $\log T_{\rm eff}= 4.187$.
The model parameters are chosen according to the discussion in \citet{muk13}.

In the upper panel, the frequency groups at $\sim\!0.8$ and $\sim\!1.6$~mHz are consistent with $m=1$ r modes for a rotation frequency of $1.75$~mHz.  
The frequency group around $2.5$~mHz may be identified as $m=2~(k=-1)$ odd modes, or 
 combination frequencies between the frequency groups of around $0.8$ and $1.6$~mHz.

Frequencies in 2010 (red lines in the lower panel of Fig.~\ref{fig:eqlyn}) are very close to frequencies obtained before the outburst, while the frequencies detected in 2012--2014 
(blue lines) are slightly higher.
So, a slightly faster rotation of $1.90$~mHz is needed to be consistent with 2012--2014 frequencies (blue lines), although
for the 2010 frequencies the pre-outburst rotation frequency should be good enough. 
This may indicate that the rotation was spun up slightly by accretion during the quiescent period between 2010 and 2012; during the period, pulsation signals were hardly detected as discussed in \citet{muk13}. 

The change in rotation frequency needed to explain pulsation frequency shifts between before and after the 2006 outburst of EQ~Lyn is much smaller than the case of GW~Lib (\S\ref{sub:gwlib}).
The difference is consistent with the fact that the $\sim\!\!5$~mag eruption of EQ~Lyn was  much less energetic (i.e. much less accretion occurred) compared with the eruption of GW~Lib with a brightning of $\sim\!\!9$~mag. 
 
\subsection{EZ Lyn (= SDSS J080434.20+510349.2)}
EZ~Lyn, identified as a CV by \citet{szk06}, is somewhat different from other systems.
EZ~Lyn underwent two dwarf nova outbursts in 2006 March \citep{pav07} and 2010 September \citep{kat12}, the interval is much shorter than typical recurrence times of decades.   
In addition, "mini-outbursts" with amplitudes of about $0.5$~mag occur at quiescence, indicating the presence of some activities in the accretion disk \citep[the long-term light curve is documented in e.g.][]{zha13}. 

\begin{figure}
\includegraphics[width=\columnwidth]{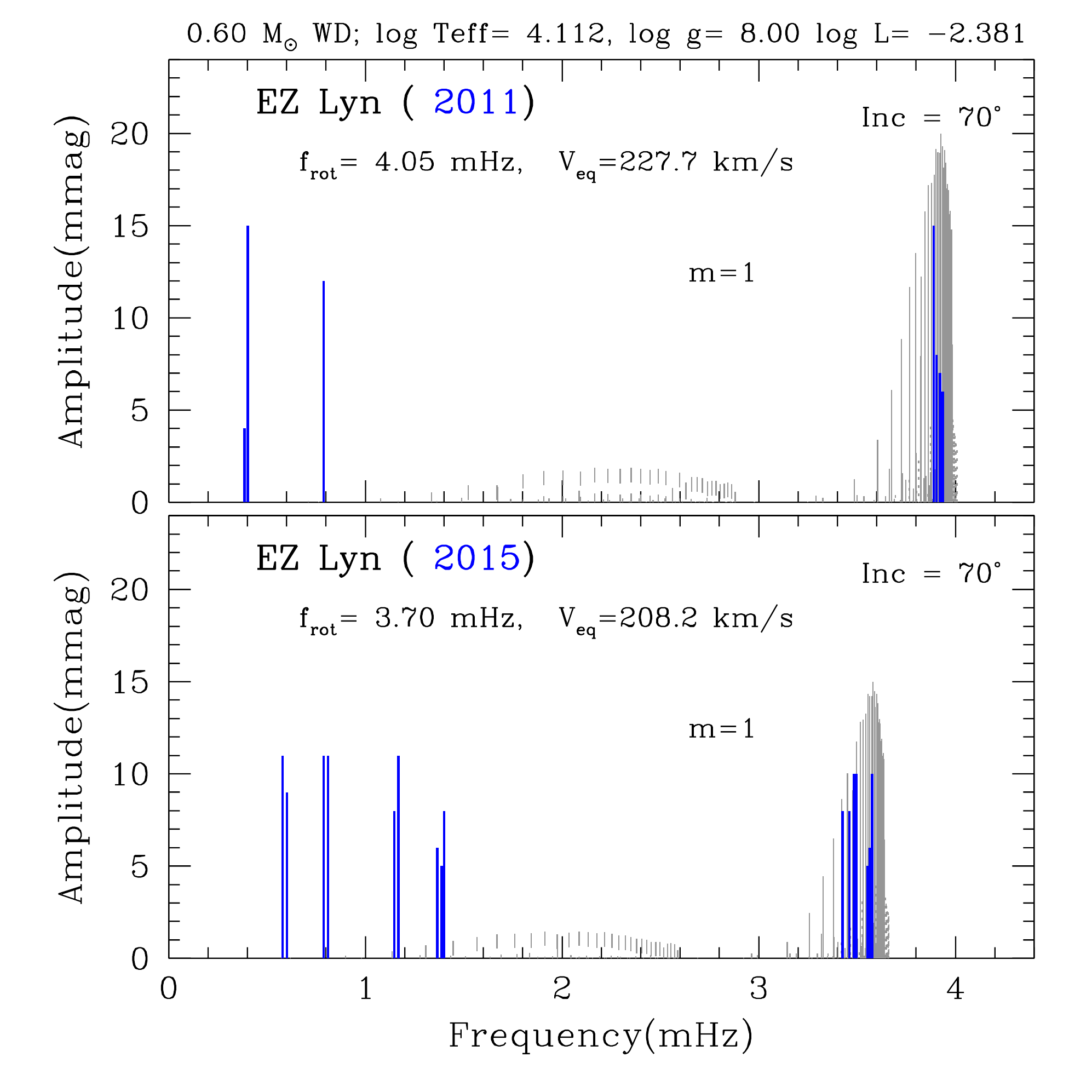}  
\caption{EZ~Lyn after 2010 ourburst.
Frequencies/amplitudes in 2011 are taken from \citet{szk13ez} and \citet{pav12}, while the data in 2015 are taken from \citet{sos17}, where the amplitudes in UV are reduced by a factor of 11 to be consistent with optical amplitudes \citep{szk13ez}.
The upper panel compares the 2011 data with r modes (gray lines) of $m=1$ for a rotation frequency ($f_{\rm rot}$) of $4.05$~mHz, while the lower panels compares the 2015 data with r modes for $f_{\rm rot}=3.70$~mHz. Because of the presence of a shallow eclipse \citep{kat09,szk13ez} a relatively high inclination is adopted.}
\label{fig:ezlyn}
\end{figure}

Prior to the 2010 outburst, a periodicity of $12.6$~min ($1.32$~mHz), which can be attributed to nonradial pulsation, was detected from 8 months to 2.5~yr after the 2006 outburst \citep{pav09,zha13}.
But after that no short periodicities were detected before the 2010 outburst.

About 7 months after the 2010 outburst \citet{pav12} found a much shorter periodicity at $4.28$~min ($3.89$~mHz). \citet{szk13ez} confirmed it 14 months after the 2010 outburst by UV observations with HST (Hubble Space Telescope) as well as ground-based optical observations.
(\citet{pav14} reported a higher frequency at $4.25$~mHz present for a week in 2012 November.)  
Furthermore, \citet{sos17} detected similar but slightly lower frequencies in 2015--2016 observations.

Fig.~\ref{fig:ezlyn} shows pulsation frequencies detected in 2011 (upper panel) and 2015 (lower panel).
Visibilities (gray lines) of r modes are shown for a $0.6~M_\odot$ WD model with $f_{\rm rot}=4.05$~mHz ($=350$~d$^{-1}$) in the upper panel and $f_{\rm rot}=3.70$~mHz ($=320$~d$^{-1}$) in the lower panel.
(Model parameters, $M/M_\odot$, $T_{\rm eff}$  in Fig.~\ref{fig:ezlyn} are chosen according to the analysis by \citet{szk13ez}; $T_{\rm eff}=13,100$~K, $\log g = 8.0$.)
These rotation frequencies are chosen to fit the main frequency groups to be consistent with $m=1$ even ($k=-2$) modes.
Since a relatively high inclination of $70^\circ$ is chosen, visibilities of odd modes (dashed lines) are low.

Lower frequency peaks in both panels cannot be explained by r modes.
In the 2011 data (upper panel), the $0.4$~mHz peak probably corresponds to a harmonic of the orbital frequency $0.196$~mHz of EZ~Lyn \citep{gae09}, while the $0.8$~mHz peak could be the harmonic of $0.4$~mHz. 
More low-frequency features are seen in the 2015 data (lower panel); among them only the $0.8$~mHz peaks are common to the 2011 data, while the $0.6$~mHz peaks may be explained as combination frequencies between $0.8$~mHz and the orbital frequency.
Frequencies at $\sim\!\!1.2$~mHz are comparable to the 12.8~min periodicity ($1.32$~mHz) seen from 8 months to 2.5~yr after the 2006 outburst \citep{pav07}, although the cause of these frequencies is not clear. (Frequencies at $\sim\!\!1.4$~mHz are probably combination frequencies between the $\sim\!\!1.2$~mHz frequencies and the orbital frequency.)
   
\citet{szk13ez} obtained a projected rotation velocity of $V\sin i= 225 \pm 75$~km~s$^{-1}$ from the UV spectra of EZ~Lyn acquired in November 2011, while our model predicts an equatorial rotation velocity of 228~km$^{-1}$ for the 2011 pulsation frequencies (upper panel). 
The presence of a shallow eclipse \citep{kat09,szk13ez} indicates a high inclination.
So, the projected velocity is consistent with the equatorial rotation velocity $228$~km~s$^{-1}$ of our r-mode pulsation model for the year 2011.

Our r-mode models interpret the shift of pulsation frequencies from $\sim\!\!3.9$~mHz (2011) to $\sim\!\!3.5$~mHz (2015) as that the rotation rate of the H-rich envelope of the WD has slowed down slightly during the time interval, due to a re-distribution of angular momentum.
Further monitoring the pulsations of EZ~Lyn would be very interesting.

\subsection{V455 And (= HS 2331+3905)}   \label{sub:v455and}
\begin{figure}
\includegraphics[width=\columnwidth]{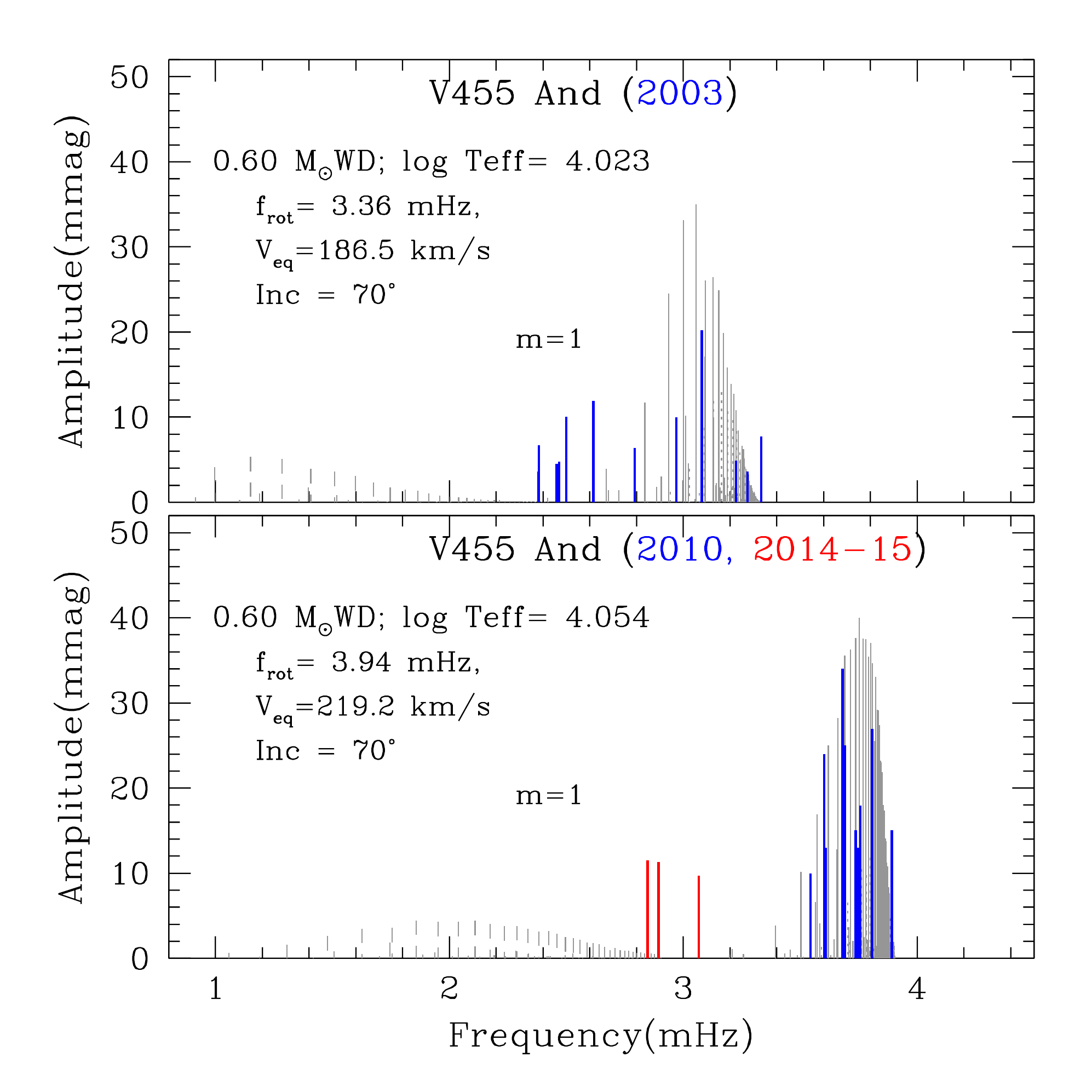}   
\caption{{\bf Upper panel:} Short period variations of V455~And before the 2007 outburst.  
Frequencies/amplitudes obtained by \citet{ara05} are compared with r modes of $m=1$ (gray lines) in a $0.6~M_\odot$ WD model with a rotation frequency ($f_{\rm rot}$) of 3.36~mHz.
{\bf Lower panel:} V455~And after the outburst. 
Frequencies in 2010 (July--September) (blue lines) 
are taken from \citet{sil12},
while 2014--2015 data (red lines) 
are taken from  \citet{muk16}. 
These data are compared with $m=1$ r-mode visibilities for $f_{\rm rot}=3.94$~mHz.}
\label{fig:v455and}
\end{figure}
The cataclysmic variable V455~And has shown complex variabilities. It underwent a large dwarf-nova outburst in September 2007 \citep{tem07v455}. 
Prior to the outburst, \citet{ara05} and \citet{gae07} found short period variations in 2002--2003 and in 2004, respectively. 
They also found an orbital period of 81.08~min with a shallow eclipse.
The short period variations consist of a group at $\sim\!\!300-400$~s and a $67.6$~s period with its harmonic.
The latter is thought to be caused by a spinning magnetic dipole (intermediate polar), and the former are attributed to pulsations in the primary accreting WD. 

After the 2007 outburst, the photometric and spectroscopic variations have been monitored by \citet{sil12}, \citet{szk13} and \citet{muk16}.
While the $67.6$~s period is almost constant \citep{muk16}, the pulsation periods have decreased to $\sim\!\!250-280$~s \citep{sil12}. 
Furthermore, \citet{szk13} found that these periods are, in turn, getting longer with time. 
Fig.\,\ref{fig:v455and} shows the pulsation frequencies/amplitudes  before (upper panel) and after (lower panel) the 2007 outburst.
As mentioned above, the pulsation frequencies got higher after the outburst, while after that they decreased considerably from 2010 (blue lines) to 2014-15 (red lines).
Gray lines are frequencies/visibilities of $m=1$ r modes in WD models of $0.6~M_\odot$ with rotation frequencies as indicated.   

\citet{ara05} determined the effective temperature of the accreting WD in V455~And at quiescence prior to the 2007 outburst as $T_{\rm eff} = 10,500$~K, while
\citet{szk13} obtained effective temperatures of $11,100\pm 250$ from 2010 HST UV spectra and $10,850\pm300$~K from 2011 spectra.
The effective temperatures of the models in Fig.~\ref{fig:v455and} are chosen taking into account above determinations, although r-mode properties are not sensitive to $T_{\rm eff}$.

In order to make r-mode frequency ranges to be roughly consistent with the observed pulsation frequencies, we choose rotation frequencies of $3.3$~mHz ($303$~s in period) for the 2003 data (upper panel) and $3.94$~mHz ($254$~s) for the 2010 data (lower panel).
Obviously, these rotation rates are much ($\sim\!\!4$ times) slower than the spin rate of the dipole magnetic poles. 
We can argue that this may indicate the presence of a considerable differential rotation between the H-rich layers and the degenerate core of the WD.
Since visible r modes should be trapped in the H-rich layers, the pulsation frequencies reflect the rotation rate there. 
On the other hand, the dipole magnetic field, which interacts with circum-stellar gas,  should be rooted in the central part of the white dwarf and hence rotate at a rate of the core rotation;i.e., the $67.6$~s period should be the core-rotation period, which can be different from the rotation period of the envelope.
Since a large inertia is expected in the core, it is reasonable that the period $67.6$~s was not changed by the outburst \citep{muk16}.

The gradual decrease of pulsation frequencies mentioned above appears in the lower panel of Fig.~\ref{fig:v455and}; the blue-line group for 2010 data is separated from the red-line group for 2014--2015 data.
The rotation frequency to fit with the 2014--2015 group should be comparable with the rotation frequency prior to the outburst, which indicates the rotation rate in the outer H-rich layers to be slowing down nearing the pre-outburst rotation rate. 

R-mode oscillations of the accreting WD in V455 And should be generated mechanically by the disturbed flows caused accretion or/and the interaction with the differentially rotating magnetic field. 
However, it still remains unsolved why the r-mode period group and dipole signal are predominantly detected in emission-line photons as found by \citet{szk13}. 

\subsection{PQ And}
\begin{figure}
\includegraphics[width=\columnwidth]{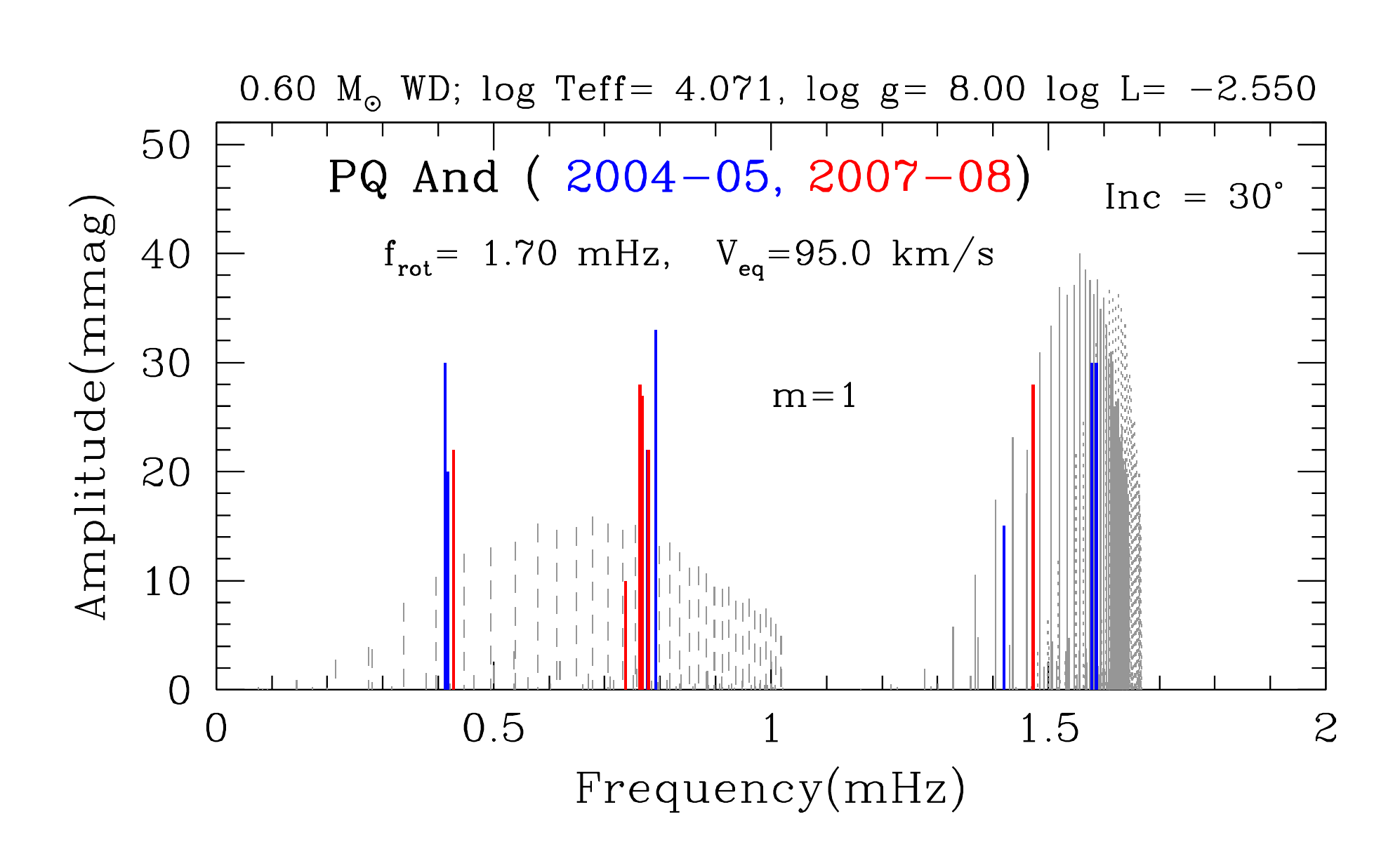}  
\caption{The visibility distribution of r modes of a $0.6~M_\odot$ WD with a rotation frequency of $1.70$~mHz is compared with short periodicities detected in PQ And.
Data for $2004-2005$ (blue) are taken from \citet{vanl05,pat05pq}, and data for $2007-2008$ (red) from \citet{szk10}. \citet{pat05pq} identified frequencies at $0.41$~mHz as the harmonic of the orbital frequency.}
\label{fig:pqand}
\end{figure}

The cataclysmic variable PQ~And underwent outbursts in 1938, 1967, and 1988.
Short-period light variations at quiescence were discovered in 2004 by \citet{vanl05}, and confirmed by \citet{pat05pq} in 2005.
\citet{szk10} detected similar periodicities in 2007--2008.
The observed frequencies/amplitudes are shown in Fig.~\ref{fig:pqand} with blue and red solid lines. These frequencies form three groups.
\citet{pat05pq} found the frequency $0.41$~mHz ($40$~min) to be the harmonic of the orbital frequency.

The other two frequency groups are compared with the visibility distribution (gray lines) of r modes for a $0.6~M_\odot$ WD model with a rotation frequency of $1.70$~mHz (606~s in period).
The rotation frequency has been chosen in order for the frequencies around $1.5$~mHz to agree with $m=1~(k=-2,-3)$ r modes.
The observed frequencies in the range $0.75-0.8$~mHz lie in the frequency range of $m=1~(k=-1)$ r modes.
The model parameters are consistent with the spectroscopic analysis by \citet{sch04}, who obtained $T_{\rm eff} = 12,000 \pm 1,000$~K and $\log g = 7.7 \pm 0.3$.
An inclination angle of $30^\circ$ is chosen arbitrarily based on no eclipse being reported.

\subsection{GY Cet (=SDSS J013132.39$-$090122.3)}
\begin{figure}
\includegraphics[width=\columnwidth]{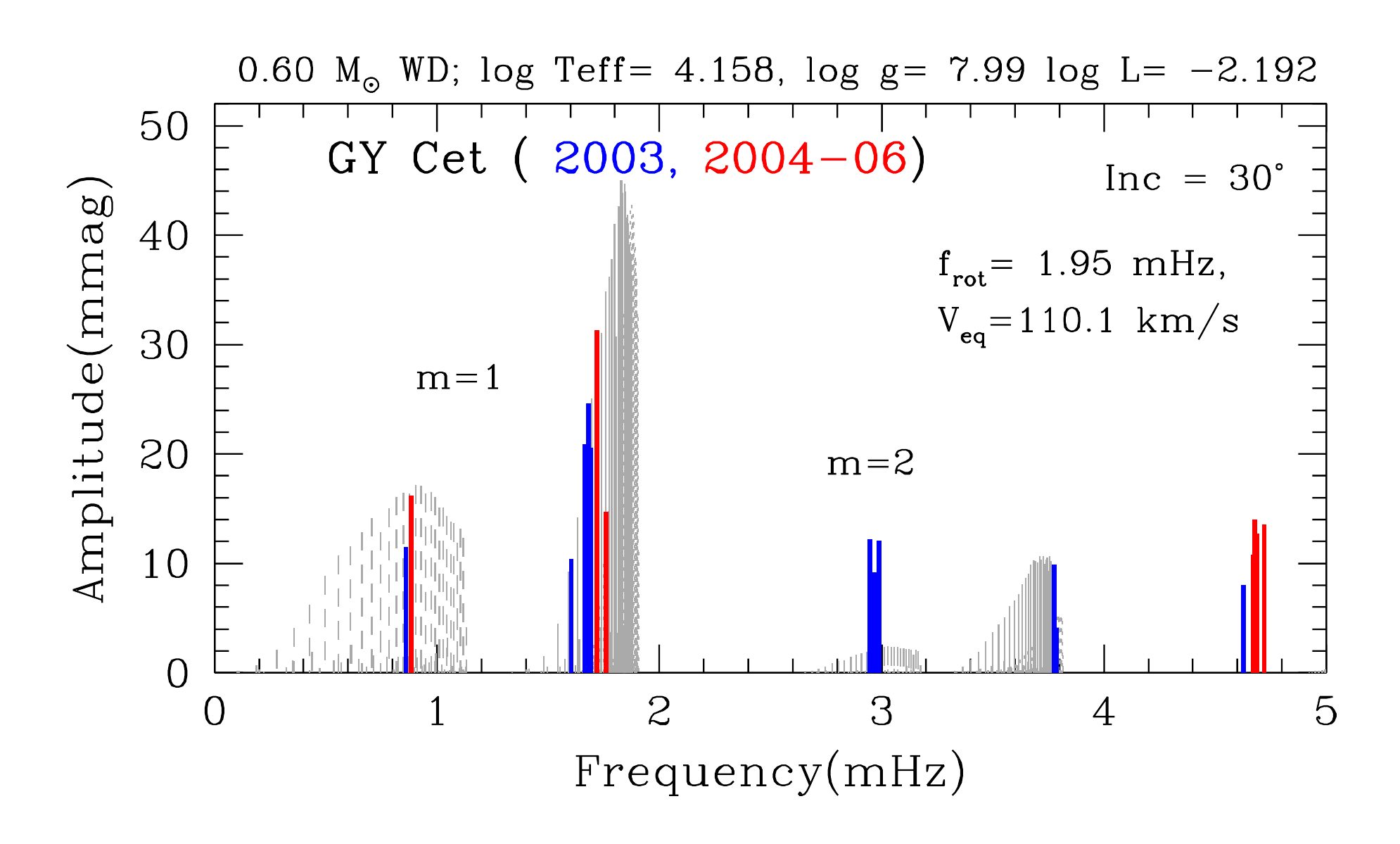}   
\caption{Predicted visibilities of r modes for a $0.6~M_\odot$ WD model with a rotation frequency of $1.95$~mHz (gray lines) are compared with pulsation frequencies/amplitudes of GY~Cet listed in \citet{szk07apj}. Each frequency/amplitude is colour-coded by the observed year as indicated. (The UV amplitude at $4.7$~mHz (2005) is reduced by a factor of 6 to be consistent with optical amplitudes.)
}
\label{fig:gycet}
\end{figure}
\citet{szk03} found GY~Cet to be one of the CV with very low mass-transfer rates. No outburst of GY~Cet has been observed. 
\citet{war04} discovered short-period light variations in 2003.
\citet{szk07apj} obtained an effective temperature of $14500$~K for the primary WD in GY~Cet. 
\citet{szk03} obtained an orbital frequency of $0.17$~mHz,
while \citet{otu16} gives an orbital inclination of $25^\circ$ for the system.

Fig.~\ref{fig:gycet} compares the observed pulsation frequencies/amplitudes listed in \citet{szk07apj} with r mode visibilities predicted for a $0.6~M_\odot$ WD model with a rotation frequency of $1.95$~mHz.
The rotation frequency is chosen in order to include as many observed frequencies as possible into the predicted range of r-mode frequencies, although the frequency group at $\sim\!\!4.7$~mHz  cannot be explained by r modes.
To fit this frequency group with r modes of $m=1$ $(k=-1,-3)$, a rotation frequency of $\sim4.80$~mHz would be needed. With this rotation frequency, however, the main frequency group at $\sim1.7$~mHz and the frequency group  at $0.9$~mHz cannot be consistent with r modes. So, we consider this choice unlikely. 
The frequency group at 4.7~mHz may correspond to 
 combination frequencies between groups at $\sim\!\!1.7$ and $\sim\!\!3.0$~mHz \citep[as indicated in table 4 of][]{szk07apj}. 
 As another possibility, 4.7~mHz could be the rotation rate of the WD core emerged by a magnetic field anchored in the core similarly to the 67~s periodicity of  V455~And (\S\ref{sub:v455and}). Needless to say, further observations are desirable; in particular, a spectroscopic determination of $V\sin i$ would be helpful to understand the periodicities of GY~Cet.

\subsection{MT Com (= RE J1255+266)}
\label{sub:mtcom}

\begin{figure}
	\includegraphics[width=\columnwidth]{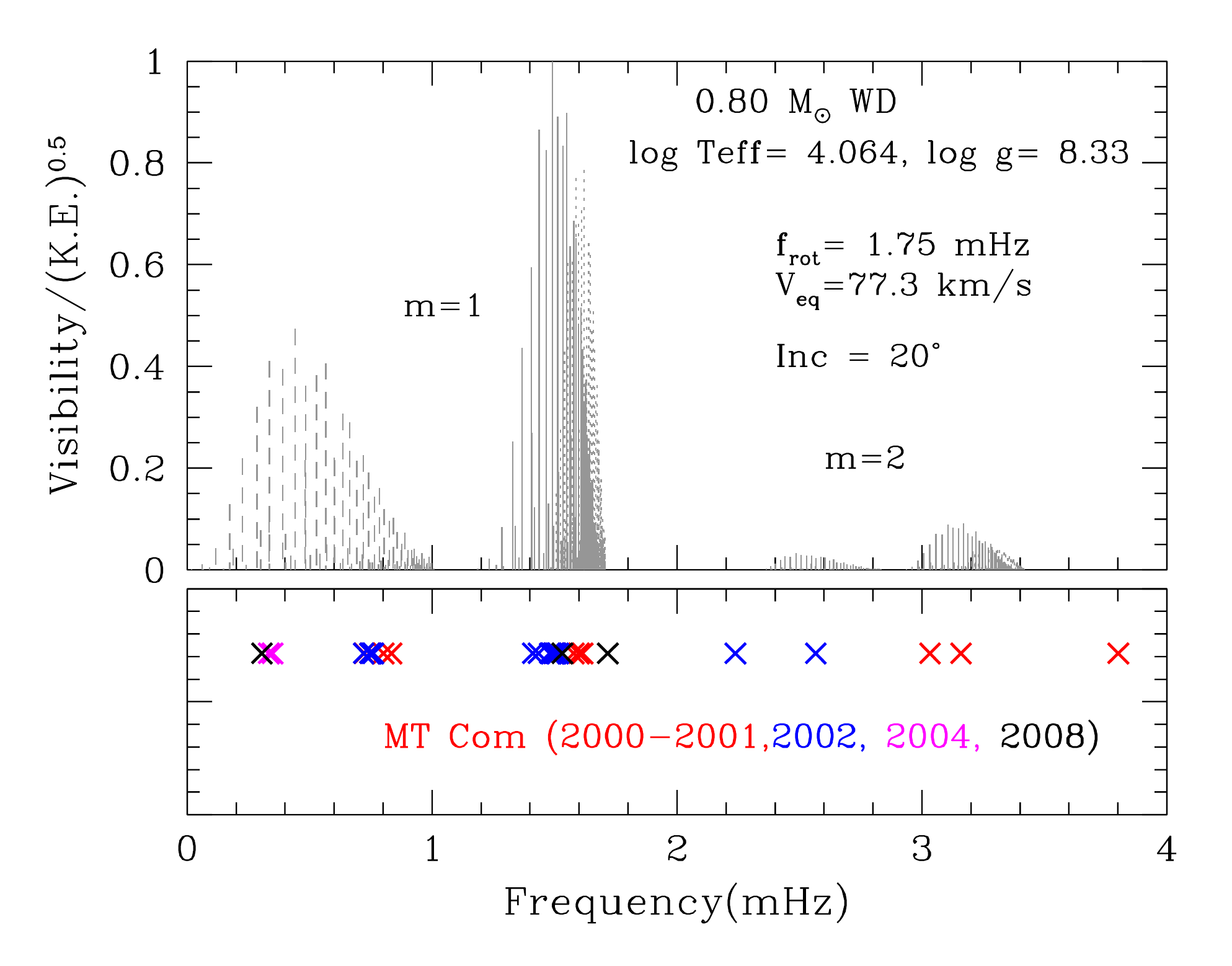}  
    \caption{Frequencies (crosses; lower panel) of short period variations detected in MT Com by \citet{pat05} (in $2000-2004$) and \citet{szk10} (in 2008) are compared with r-mode frequency ranges (gray lines) for a $0.8M_\odot$ WD model with a rotation frequency of 1.75 mHz (upper panel). 
(Two very high frequencies of 8.29 and 8.37~mHz detected by \citet{pat05} are not shown.) Observational years are colour-coded as indicated. (The vertical axis in the upper panel is arbitrarily normalized.)}
    \label{fig:mtcom}
\end{figure}

MT Com was discovered as a bright Extreme Ultraviolet (EUV) transient in 1994 with ROSAT \citep{dah95}, which is regarded as a normal dwarf nova outburst \citep{whe00}. From optical time-series photometry in 2000--2004,  \citet{pat05} detected many short period signals mainly around 1344, 1236, and 668~s (0.744, 0.809, and 1.497~mHz, respectively).
These periodicities as well as those obtained by \citet{szk10} are shown in the lower panel of Fig.~\ref{fig:mtcom}. 
The orbital period  $0.0829$~d ($0.140$~mHz) \citep{pat05} is not shown.

\citet{pat05} estimated $T_{\rm eff}=13,000\pm2000$~K for the primary WD in MT~Com, while \citet{szk10} obtained $12,000\pm1000$~K.
\citet{pat05} also argued the primary white dwarf in MT~Com should be fairly massive and the inclination should be  $i \sim\!\!20^\circ$ if the mass ratio is as low as $\sim\!\!0.03$.
Our model parameters are chosen by taking into account these results.

The upper panel of Fig.~\ref{fig:mtcom} presents visibility of r modes predicted for a WD model of $0.8~M_\odot$ with a rotation frequency of $1.75$~mHz ($9.5$~min).
The rotation frequency is chosen such that r-mode frequencies should be consistent with as many observed frequencies as possible.
Fig.~\ref{fig:mtcom} shows that r modes are roughly sucessful in explaining most short periodicities observed in MT~Com.

\subsection{BW Scl (=2MASS J23530086$-$3851465)}
\begin{figure}
\includegraphics[width=\columnwidth]{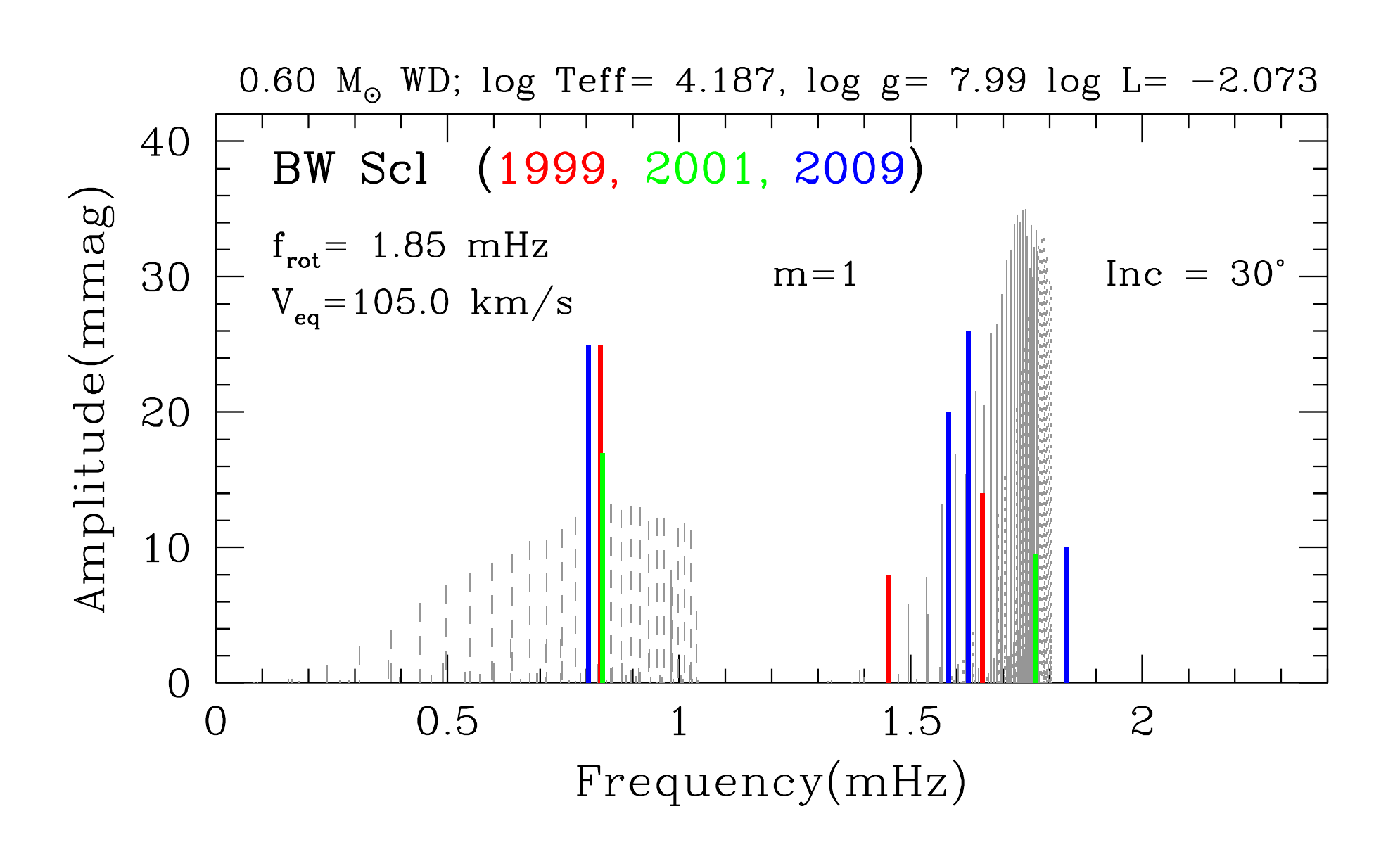}   
\caption{Frequencies of short period variations of BW~Scl obtained by \citet{uth12} in 1999 (red), 2001 (green), and 2009 (blue) are shown, while combination frequencies with the orbital frequency and its harmonic are removed. (Amplitudes are taken from table 2 and Fourier spectra for each observing campaign of \citet{uth12}).
Gray lines show visibilities of $m=1$ r modes in a $0.6~M_\odot$ WD with a rotation frequency of $1.85$~mHz.
Since no information on the inclination of BW~Scl is available, a relatively low inclination, $30^\circ$, is chosen arbitrarily.
}
\label{fig:bwscl}
\end{figure}
\citet{abb97} identified BW~Scl as a CV.
\citet{uth12} discovered short period light variations in BW~Scl at quiescence. 
The short-period (10 to 20 min) variations are attributed to nonradial pulsations of the accreting primary WD.
\citet{uth12} reported modulations in the pulsation frequencies and amplitudes, suggesting the presence of more complex frequency structure than detected.

Fig.~\ref{fig:bwscl} compares the detected pulsation frequencies with r-mode visibilities calculated for a $0.6~M_\odot$ WD model assuming a rotation frequency of $1.85$~mHz.
Model parameters are chosen according to the spectroscopic analysis by \citet{gae05}, who obtained $T_{\rm eff} = 14,800\pm900$ (with assumed $\log g = 8.0\pm0.5$). 

A group of observed frequencies between $1.4 - 1.7$~mHz ($12-9.8$~min in periods) is fitted to $m=1$ even ($k=-2$) or odd ($k=-3$) r modes, while frequencies around $0.8$~mHz (20~min in period) to odd ($k=-1$) r-modes.  
Predicted equatorial rotation velocity, $105$~km~s$^{-1}$,  is consistent with the upper bound $V\sin i \le 300$~km~s$^{-1}$ obtained by \citet{gae05} for the primary WD in  BW~Scl.

The first outburst of BW~Scl was detected by M.Linnolt in 2011 October as mentioned in \citet{kat13}.
So far, no information on short-period variations after the outburst is available in the literature.

\subsection{V386 Ser (= SDSS J161033.64$-$010223.3)}
\begin{figure}
\includegraphics[width=\columnwidth]{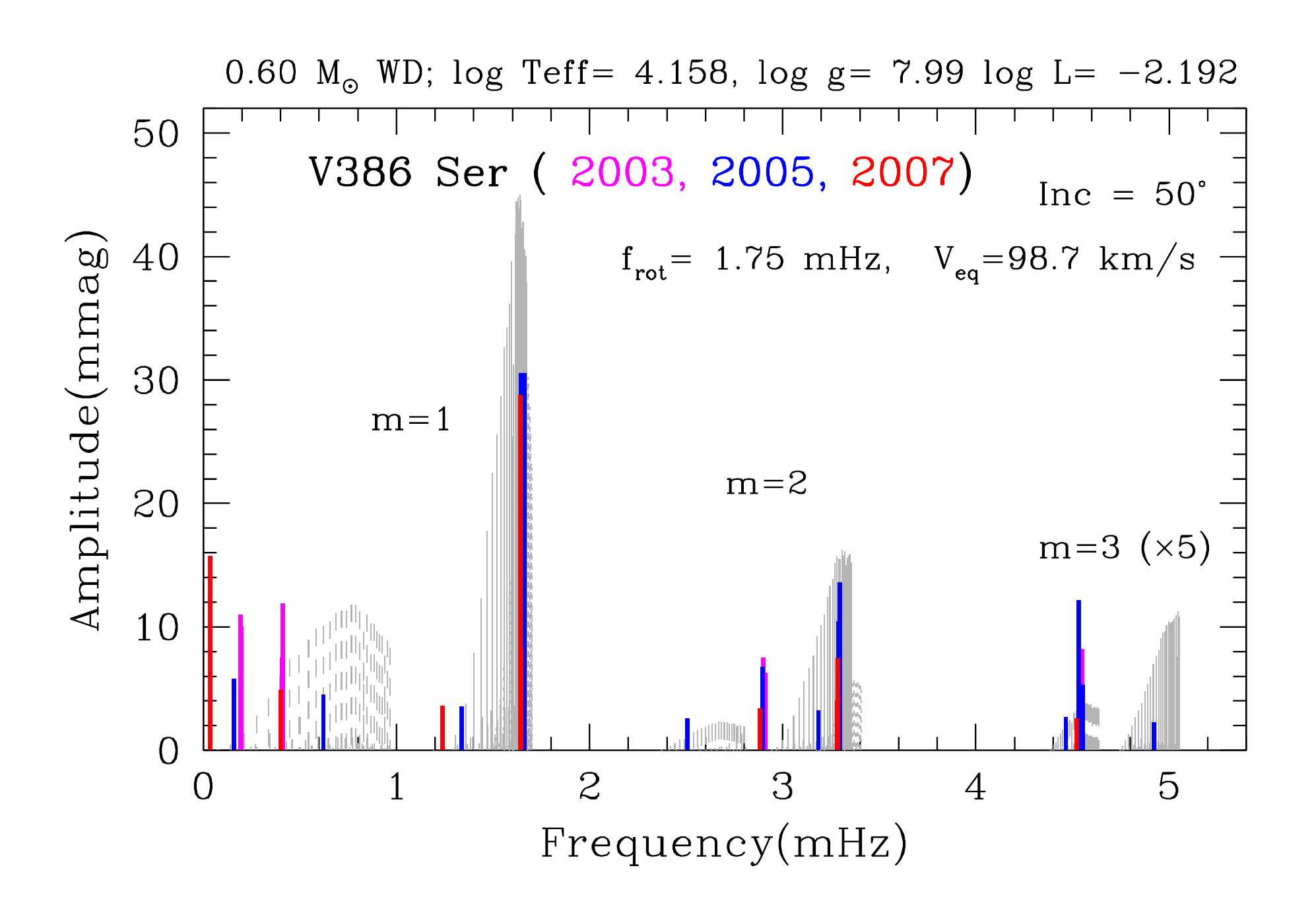}  
\caption{Observed frequencies of V386~Ser in 2003 by \citet{wou04} (magenta), in 2005 by \citet{cop09,szk07apj} (blue), and in 2007 by \citet{muk10} (red) are compared with visibility distribution of r modes for a $0.6~M_\odot$ WD model with a rotation frequency of $1.75$~mHz.
The predicted visibilities of $m=3$ r modes are multiplied by a factor 5, to make the frequency range more apparent. }
\label{fig:ser}
\end{figure}
V386 Ser is one of the CVs selected from SDSS spectra by \citet{szk02}.
No outburst of V386~Ser has been detected.  
Short period oscillations of the accreting WD in V386~Ser were discovered, next to GW Lib, by \citet{wou04} in 2003. 
The short periodicities were also observed by \citet{cop09} in 2005, and \citet{muk10} in 2007. 
The observed frequencies are similar but not the same among the three seasons of  observations. 

These pulsation frequencies are shown in Fig.~\ref{fig:ser}, and are compared with the visibility distribution of r modes for a $0.6~M_\odot$ WD model with a rotation frequency of $1.75$~mHz ($571$~s in period). 
The effective temperature of the model is chosen according to \citet{szk10}.
The rotation frequency is chosen for the main frequencies at $1.64$~mHz to be consistent with r modes of $m=1 (k = -1, {\rm or} -3)$.

Fig.~\ref{fig:ser} does not include frequencies around $6$~mHz detected by \citet{wou04} and \citet{cop09} because they are considered to be combination frequencies \citep{wou04,cop09}, and those frequencies were not detected in the multi-site observations in 2007 as reported in \citet{muk10}.
In the low frequency region, the orbital frequency at $0.2$~mHz and its harmonic at $0.4$~mHz have appreciable amplitudes, which cause various combination frequencies.
\citet{muk10} identified the frequency at $2.9$~mHz is a combination between the harmonic of the main frequency at $1.64$~mHz and the harmonic of the orbital frequency $0.2$~mHz (i.e., $(1.64-0.2)\times 2 = 2.88$).

\citet{muk10} found that the main frequency at $1.64$~mHz is an equally spaced triplet.
Assuming the triplet as rotational splitting, they derived a rotation period of 4.8~d ($0.002$~mHz), which is much slower than our rotation rate assumed ($1.75$~mHz) for the model (the latter is more typical as an accreting white dwarf).
The discrepancy would be resolved if the triplet splitting $\sim\!\!0.001$~mHz is caused by    narrow frequency spacings among adjacent r-mode frequencies; the frequency spacing is in fact comparable to $\sim\!\!0.001$~mHz at $\sim\!\!1.64$~mHz for the $m=1$ r modes shown in Fig.~\ref{fig:ser}.
In addition, beatings among dense r-mode frequencies may explain the small frequency differences in different observing runs as shown in Figs.7~and~8 of \citet{muk10}. 
The dense r-mode frequency spectrum is also consistent with the amplitude variations found and inferred as unresolved multiplets by \citet{wou04}.

\subsection{V355 UMa (= SDSS J133941.11+484727.5)}
\begin{figure}
\includegraphics[width=\columnwidth]{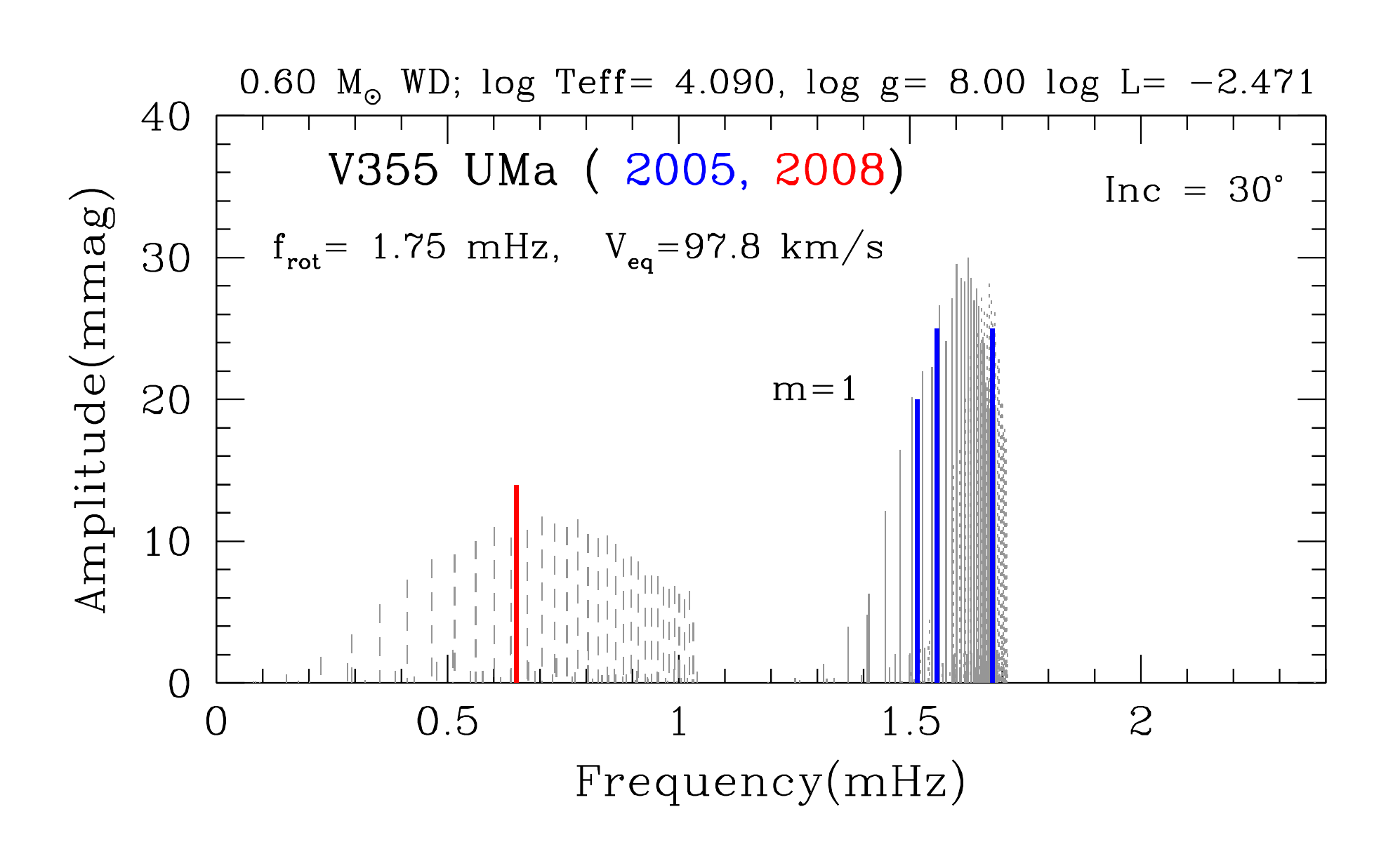}  
\caption{R-mode visibilities calculated for a $0.6~M_\odot$ WD models with a rotation frequency of $1.75$~mHz are compared with observed pulsation frequencies in V355~UMa. \citet{gae06} obtained a frequency of $1.558$~mHz in 2005 April, \citet{nil06} obtained frequencies of $1.517$ and $1.678$~mHz in 2005 August, while \citet{szk10} obtained a frequency of $0.650$~mHz in 2008 January.}
\label{fig:v355uma}
\end{figure}
V355~UMa was identified as a CV from the SDSS spectrum by \citet{szk05}.
\citet{gae06} and \citet{nil06} independently discovered, in 2005, rapid periodic signals at 600--660~s attributable to non-radial pulsations of the primary WD in V335 UMa. In 2008 \citet{szk10} obtained a somewhat longer period at 1539~s.
As shown in Fig.~\ref{fig:v355uma},  these periods are consistent with r modes in a $0.6~M_\odot$ WD model with a rotation frequency of $1.75$~mHz (571~s).  
The effective temperature of the model is consistent with the spectroscopic analysis by \citet{gae06}; they obtained $T_{\rm eff} = 12500$~K for a fixed $\log g = 8$.

The outburst of V355~UMa was detected for the first time in 2011 February by J. Shears \citep{kat12}. 
But so far, no information on short period variations after the outburst is available in the literature.

\subsection{SDSS J1457 (= SDSS J145758.21+514807.9)}
\begin{figure}
\includegraphics[width=\columnwidth]{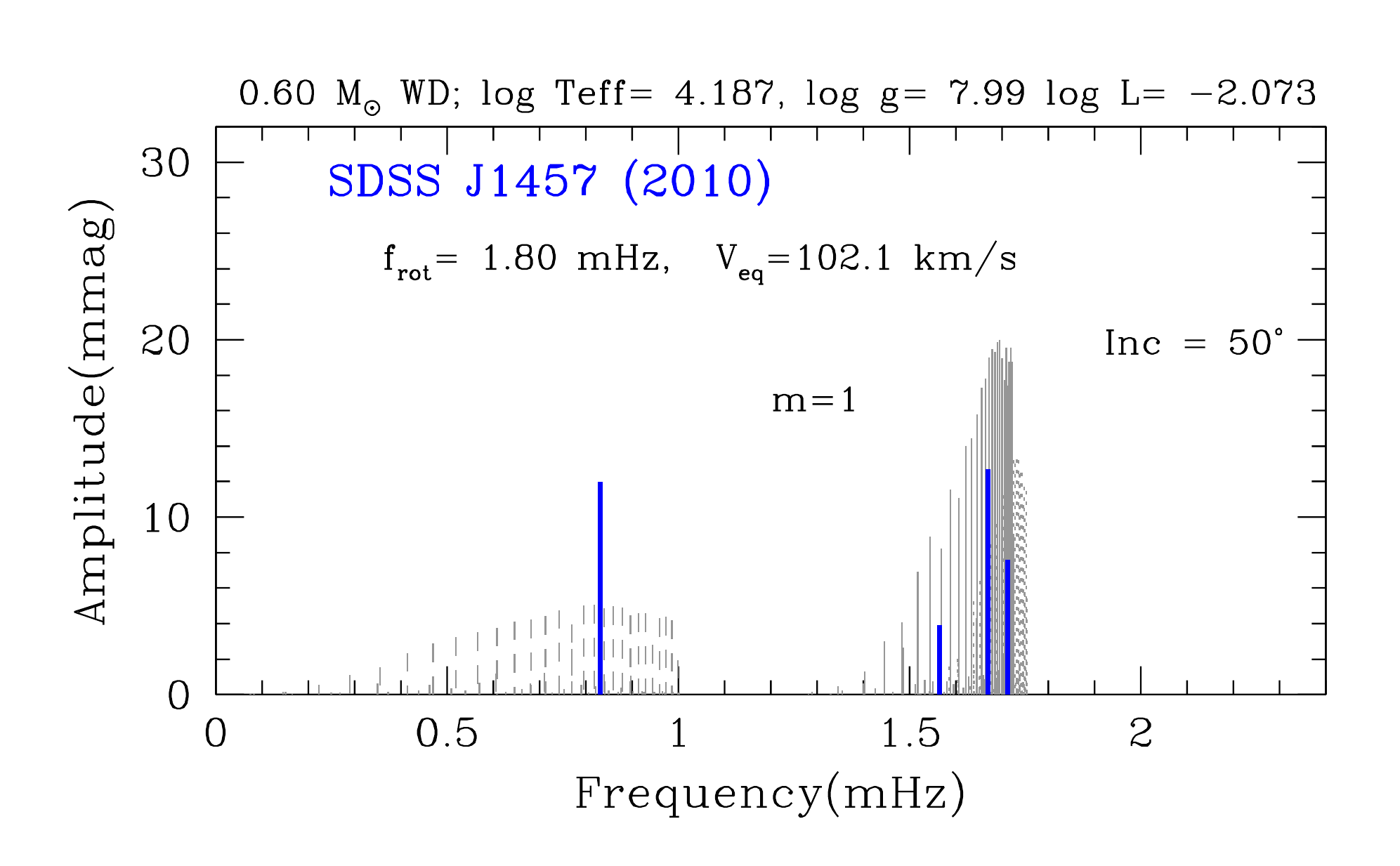}   
\caption{Frequencies and amplitudes of SDSS~J1457 obtained by \citet{uth12} are compared with visibility of r modes of $m=1$ for a $0.6~M_\odot$ WD model with a rotation frequency of $1.80$~mHz.}
\label{fig:J1457}
\end{figure}
SDSS~J1457 was identified as a CV from the SDSS spectrum by \citet{szk05}.
\citet{uth12} discovered short periodicities in SDSS~J1457 and a orbital frequency of $0.214$~mHz ($77.9$~min). 
Fig.~\ref{fig:J1457} compares these pulsation frequencies/amplitudes with r-mode visibilities calculated for a $0.6~M_\odot$ WD model with a rotation frequency of $1.80$~mHz ($556$~s).
Since no estimates of $T_{\rm eff}$ and $\log g$ are available in the literature, the same equilibrium model as that for BW~Scl (which has similar pulsation frequencies) is adopted.
Fig.~\ref{fig:J1457} shows the observed short periodicities in SDSS~J1457 consistent with r modes of $m=1$. 

\subsection{SDSS J2205 (= SDSS J220553.98+115553.7)}
\begin{figure}
\includegraphics[width=\columnwidth]{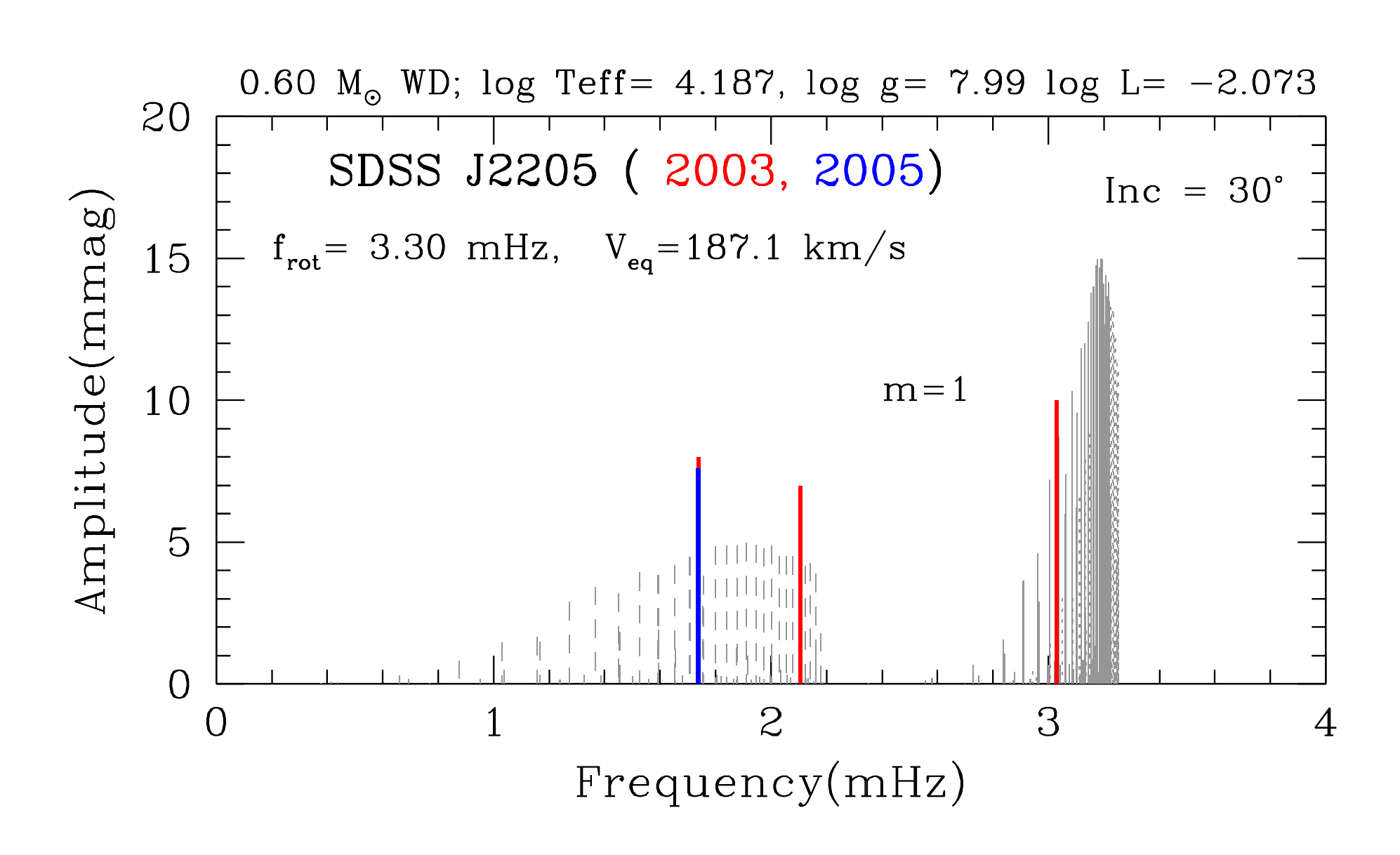}  
\caption{Pulsation frequencies of SDSS~J2205 obtained so far are compared with visibility distribution of r modes in a $0.6~M_\odot$ WD model with a rotation frequency of $3.30$~mHz.
Amplitude at $1.736$~mHz obtained by UV observations with HST \citep{szk07apj} is reduced by a factor of $6$ to be consistent with optical observations.  }
\label{fig:j2205}
\end{figure}
SDSS~J2205 was identified as a CV from the SDSS spectrum by \citep{szk03}.
\citet{war04} discovered $330$, $475$, $575$~s periodicities in 2003, while
\citet{szk07apj} confirmed the $575$~s periodicity from UV observations with HST. 
These periodicities are consistent with $m=1$ r modes predicted for a rotation frequency of $3.30$~mHz (303~s) as shown in Fig.~\ref{fig:j2205}.
The effective temperature of the model is chosen to be consistent to the value $15,000$~K obtained by \citet{szk07apj}. 

However, \citet{sou08} found in August 2007 that SDSS~J2205 had stopped pulsating. 
Since, r modes in an accreting white dwarf are probably generated by flow disturbances on the stellar surface due to accretion, diminishing r mode oscillations might mean that accretion on the white dwarf had stopped or had been very weak for a long time in SDSS~J2205.  

An dwarf-nova outburst of SDSS~J2205 was, for the first time, detected in 2011 May by CRTS  as noted in \citet{kat13}.
So far, no information on short periodicities in SDSS~J2205 after the outburst are available in the literature.

\section{Discussion}   \label{sec:discussion}
In the previous section, we have compared pulsation frequencies detected in accreting white dwarfs of twelve CVs with expected frequency ranges of r modes, determining the best rotation rate for each case. 
(Additional five stars for which comparison are less certain are discussed in Appendix.)
Table~\ref{tab:sum} gives thus obtained rotation periods and equatorial rotation velocities, $V_{\rm eq}$ as well as model $T_{\rm eff}$ and masses. 
Since visible r modes are trapped in the H-rich layers,  we regard the rotation periods as representing the rotation of the outermost layers, so that these values could be 
substantially different from the rotation rates in the WD core.

A WD mass of $0.6~M_\odot$ is adopted for most cases except for GW~Lib, MT~Com, and OV~Boo  ($0.8~M_\odot$) for which higher masses are indicated in the literature.
R-mode frequencies of a model for a rotation frequency are insensitive to the adopted mass, hence the rotation frequency determined by comparison with observed pulsation frequencies is insensitive to the mass.
However, equatorial velocity is lower for a massive white dwarf for a given rotation frequency because of the smaller radius.

\begin{table}
	\begin{center}
	\caption{Rotation rates of H-rich layers and model parameters}
	\label{tab:sum}
	\begin{tabular}{lccccc} 
		\hline
		Name  & $P_{\rm orb}$ & $P_{\rm rot}$ & $V_{\rm eq}$ & $q^{\rm c}$ & $(\log T_{\rm eff}$, $M$)\\
		      &  (hr)  & (s)   & (km\,s$^{-1}$)  \\
		\hline
		OV\,Boo$^*$           & 1.11 & 454 & 98 && (4.156, 0.8) \\
		GW\,Lib(before OB$^{\rm a}$) & 1.28  & 351 & 128 & 0.069 &(4.199, 0.8)\\
		GW\,Lib(after OB) & 1.28  & 270 & 167 & 0.069 &(4.250, 0.8)\\
	    BW Scl           & 1.30 & 541 & 105 & 0.067 &(4.187, 0.6)\\
	    SDSS J1457        & 1.30 & 556 & 102 & &(4.187, 0.6)\\
		EQ\,Lyn (before OB) & 1.32 & 571 & 99  &  &(4.187, 0.6)\\
		EQ\,Lyn (after OB) & 1.32 & 526 & 108 &  &(4.187, 0.6) \\
		V455 And (2003)$^{\rm b}$& 1.35 & 298 & 187 & 0.080 &(4.023, 0.6)\\
		V455 And (2010)$^{\rm b}$& 1.35 & 254 & 219 & 0.080 &(4.054, 0.6)\\
	    PQ And           & 1.34 & 588 & 95 &  &(4.071, 0.6)\\
        LV Cnc$^*$           & 1.36 & 250 & 225 & &(4.134, 0.6) \\
	    V355 UMa         & 1.38 & 571 & 98 & 0.066  &(4.090, 0.6)\\
	    SDSS J2205        & 1.38 & 303 & 187 & &(4.187, 0.6)\\
	    V386 Ser         & 1.40 & 571 & 99 &   &(4.158, 0.6)\\
        SDSS J0755$^*$    & 1.41 & 244 & 232 &  & (4.187, 0.6)\\
		EZ Lyn (2011)    & 1.42 & 247 & 228 & 0.078 &(4.112, 0.6)\\
		EZ Lyn (2015)    & 1.42 & 270 & 208 & 0.078 &(4.112, 0.6)\\
        DY CMi$^*$       & 1.43 & 235 & 240 &  & (4.158, 0.6) \\
        PP Boo$^*$       & 1.48 & 500 & 111 &   &(4.009, 0.6)\\
	    GY Cet           & 1.63 & 513 & 110 &  &(4.158, 0.6)\\
	    MT Com           & 1.99 & 571 & 77 &  &(4.064, 0.8)\\
		\hline
	\end{tabular}
	\end{center}
	$^*$ Rotation period may have considerable uncertainty because of a small number of observed frequencies (see Appendix). \\ 
	$^{\rm a}$ OB means `dwarf-nova outburst'.\\
	$^{\rm b}$ Rotation period associated with a magnetic field, $67.6$~s, stays constant \citep{muk16}\\
	$^{\rm c}$ Mass ratio from \citet{kat15}.\\
\end{table}

The rotation rates obtained for the accreting WDs are fairly fast, although the surface equatorial velocities are, in all cases, less than $\sim\!\!8$ percent of the Kepler velocity. 
The rotation periods given in Table~\ref{tab:sum} are much shorter than those of the  single ZZ Ceti stars which are longer than 1~hr \citep[longer than 10~hr for most stars; e.g.,][]{fon08,kaw15,her17}.
A rapid rotation of the primary WD in a CV is expected because it has accreted matter as well as angular momentum from the accretion disk.

The rotation velocities given in Table~\ref{tab:sum}, however, tend to be smaller than other primary WDs in CVs (in which no pulsations have been detected); see Table~2 of \citet{sio12}.
The difference is probably related to the fact that pulsations in the primary WDs are detected only in CV systems with extremely small mass-transfer rates.

\begin{figure}
\includegraphics[width=\columnwidth]{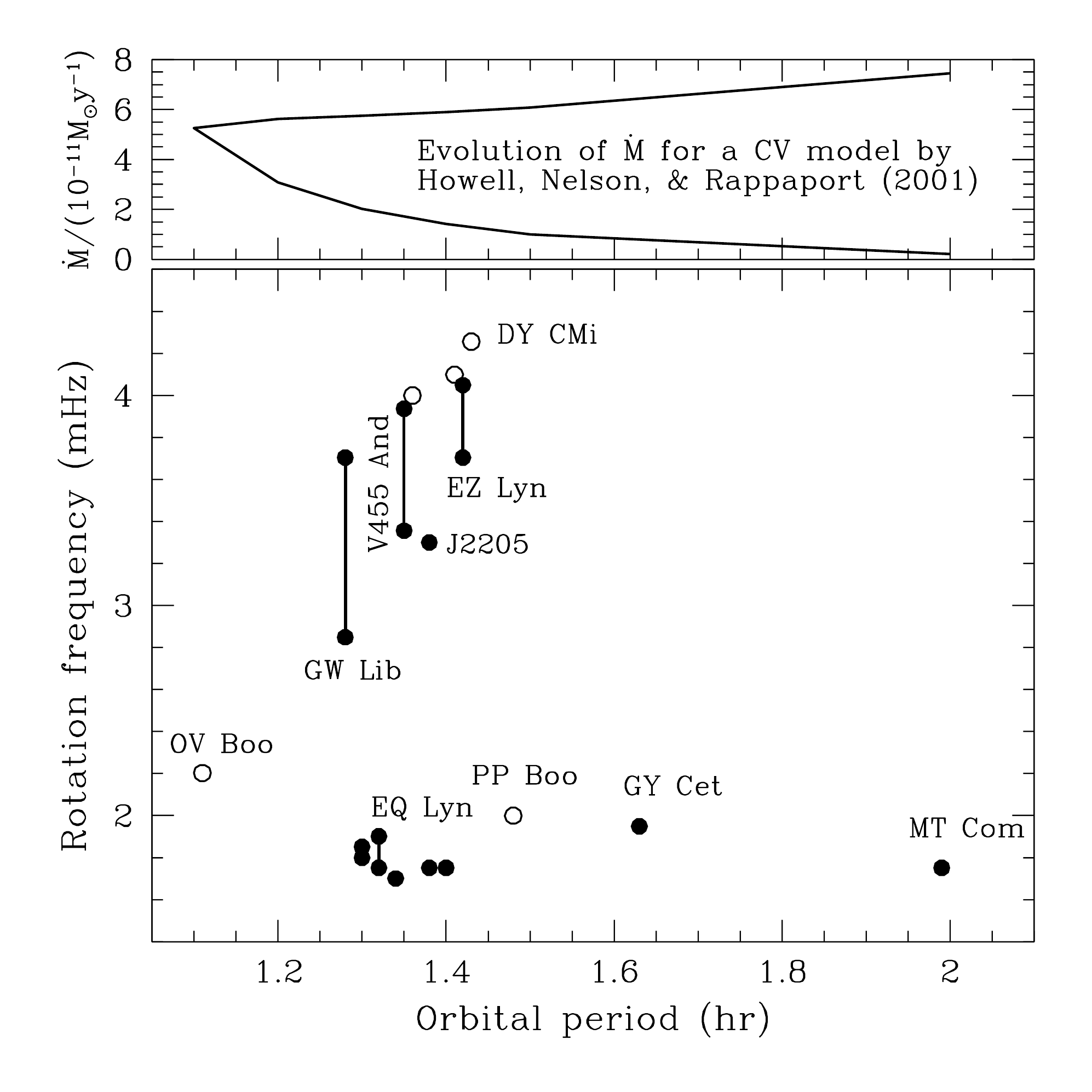}  
\caption{{\bf Lower panel:} Rotation frequency (in mHz) of primary white dwarfs obtained by comparing r mode frequencies with observed pulsation frequencies are plotted with respect to the orbital periods. 
Open circles are for less certain cases discussed in Appendix.
{\bf Upper panel:} Evolution of mass-transfer rate in a cataclysmic binary model, taken from Fig.2a of \citet{how01}.
}
\label{fig:porb}
\end{figure}
The lower panel of Fig.~\ref{fig:porb} shows rotation frequencies of accreting WDs with respect to the  orbital periods ($P_{\rm orb}$). 
(Rotation rates of the same star at different epochs are connected by a vertical line.)
This figure shows the presence of a sequence, runs from upper-right towards shorter orbital period, bounced at $P_{\rm orb}\approx 1.1$~hr, and  goes toward lower right.
For comparison, the upper panel of Fig.~\ref{fig:porb} shows a theoretical evolutionary track of the mass transfer rate ($\dot{M}$) computed by \citet{how01} for a cataclysmic binary model. 
The evolution of $\dot{M}$ starts from upper-right, comes down to the minimum orbital period at $P_{\rm orb}=1.1$~hr, and decreases to longer orbital period (lower-right).
The evolution is caused by losing the orbital angular momentum due to gravitational radiation, and by decreasing mass of the secondary star.
The shape of the sequence is similar to our sequence of rotation frequencies in the lower panel.

The similarity of the $f_{\rm rot}$--$P_{\rm orb}$ sequence to the $\dot{M}$--$P_{\rm orb}$ evolution track seems to indicate that the rotation of an accreting primary WD slows down with decreasing mass-transfer rate, rather than monotonically being spun up by accretion.
This means that some mechanism of re-distributing angular momentum in the WD works effectively.
This is consistent with the fact we found in the previous section that the rotation frequency needed to fit observed pulsation frequencies with r modes sometimes decreases after a sharp rise at an outburst as in the cases of  GW~Lib (Fig.~\ref{fig:gwlib}), EZ~Lyn (Fig.~\ref{fig:ezlyn}), and V455~And (Fig.~\ref{fig:v455and}).

The binary evolution \citep[upper panel of Fig.~\ref{fig:porb};][]{how01} predicts CVs on the lower branch (sometimes called `period bouncers') to have smaller (secondary to primary) mass ratios ($q$) compared to the cases on the upper branch at a same orbital period.
According to the mass ratios from \citet{kat15} (listed in Table~\ref{tab:sum}),
systems with lower mass ratios tend to fall on the slower rotation branch in Fig.~\ref{fig:porb}.
In addition, \citet{pat05} estimated $q < 0.06$ for MT~Com, which supports the tendency of lower $q$ for the slower rotation branch.
(\citet{pat05} also estimated a mass-transfer rate of $4\times10^{-12}M_\odot$~yr$^{-1}$ for MT~Com, which agrees with the evolutionary model.)

Thus, our $f_{\rm rot}$--$P_{\rm orb}$ sequence is likely related to the  evolution of cataclysmic binaries, and the property $f_{\rm rot}\!\sim\!2$~mHz may be considered as an additional character for the period bouncers.
Among the CVs discussed in this paper, \citet{pat11} lists MT~Com, V455~And, GW~Lib, PQ~And, BW~Scl, EZ~Lyn, and PP~Boo as period-bouncer candidates.
Based on our rotation rates, we confirm MT~Com, PQ~And, BW~Scl, and PP~Boo as period bouncers.  

\section{Conclusions}
Short-period light variations detected in accreting WDs in cataclysmic variables are identified as r-mode oscillations (global Rossby waves) that are trapped in the H-rich layers of the WD.
Since the r-mode frequencies are mainly determined by rotation frequency, we can determine a rotation frequency by making predicted r-mode frequency range to be consistent with observed pulsation frequencies for each case.
The rotation frequency should be regarded as representing the rotation speed of the outermost layers.
Thus determined rotation speeds are found to be much faster than those of non-interactive pulsating WDs (ZZ Ceti variables), but tend to be somewhat slower than the non-pulsating accreting WDs in other cataclysmic variables.

We have found that on the $P_{\rm orb}-f_{\rm rot}$ plane,  accreting WDs in cataclysmic variables lie on a sequence, which is likely related with the evolutionary sequence of mass-transfer rate as a function of orbital period for a cataclysmic binary model computed by \citet{how01}.
 
The pulsation frequencies of the primary white dwarfs in GW~Lib, EQ~Lyn, and V455~And indicate that rotation in their outermost layers spun up during a dwarf-nova outburst, while  in some cases rotation gradually slows down after the spin up.
Such shifts of rotation rates and the presence of a sequence of accreting WDs in the $P_{\rm orb}-f_{\rm rot}$ plane suggest that the rotation rate of the outermost layers of an accreting WDs does not increase monotonically with accretion; rather, it seems to settle at the `equilibrium' rate corresponding to each accretion rate. 
This suggests the presence of an efficient mechanism to transport the angular momentum in accreting WDs. 

Monitoring pulsation frequencies and spectroscopic $V\sin i$ of accreting WDs in cataclysmic variables would be very useful to understand the angular momentum accretion onto the surface and its re-distribution.

\section*{Acknowledgements}
The author is very grateful to Paula Szkody and JJ Hermes for helpful comments on a draft of this paper, also to an anonymous referee for helpful comments.




\bibliographystyle{mnras}
\bibliography{ref_pap}



\appendix

\section{R-mode fittings to observed pulsations; less certain cases }
This appendix, as in \S\ref{sec:cv}, fit r-mode oscillations  to pulsations observed in accreting WDs in cataclysmic variables, but for the cases where the fittings are less certain.

\subsection{LV~Cnc (= SDSS J091945.10+085710.0)}
\begin{figure}
\includegraphics[width=\columnwidth]{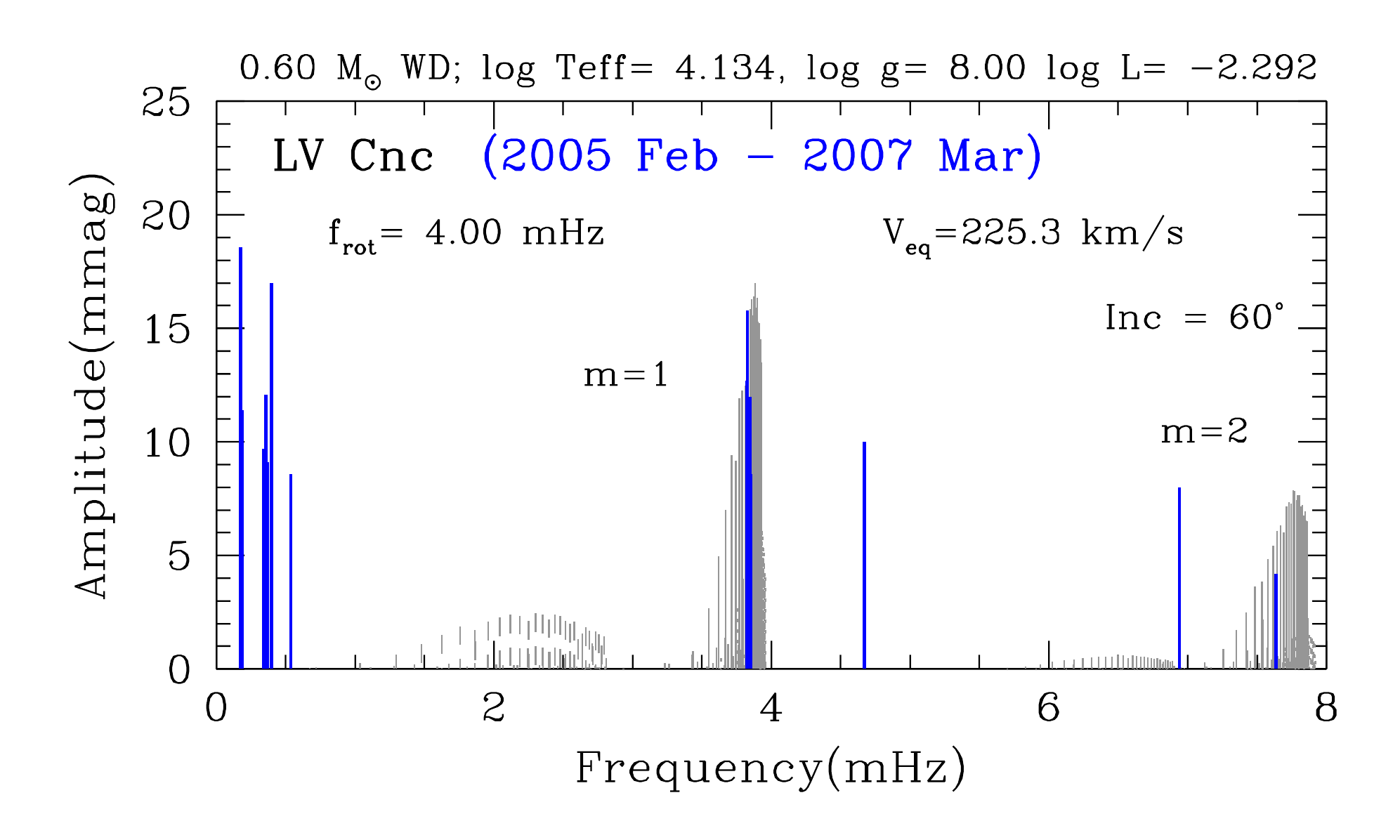}  
\caption{Frequencies detected in LV~Cnc are compared with visibilities of r modes for a $0.6~M_\odot$ WD model with a rotation frequency of $4.0$~mHz. The effective temperature is adopted from \citet{szk10}. Low frequencies less than $\sim\!\!0.6$~mHz correspond to the orbital frequency and its first and second harmonics.}
\label{fig:lvcnc}
\end{figure}
LV~Cnc was identified as a CV from the SDSS spectrum by \citet{szk05}.
\citet{muk07} discovered, in 2005 December, pulsation periods of $\sim\!\!260$~s.  
\citet{wou12} confirmed the presence of similar periodicities.
Although these short period variations had been observed until 2007 March, they were not present in 2007 November and 2008 December \citep{szk10}; i.e., the star seems to stop pulsating.

Fig.~\ref{fig:lvcnc} shows pulsation frequencies obtained by \citet{muk07} and \citet{wou12}, including low frequencies related to the orbital frequency $0.204$~mHz ($1.36$~hr) \citep{dil08} and its first and second harmonics.  
Observed $\sim\!\!260$~s pulsations ($\sim\!\!3.8$~mHz) are fitted with $m=1$ r modes for a rotation frequency of $4.0$~mHz (250~s), while frequencies at $4.7$ and $6.9$~mHz cannot be explained by r modes of the model.
 
\subsection{PP~Boo (= SDSS J151413.72+454911.9)}
\begin{figure}
\includegraphics[width=\columnwidth]{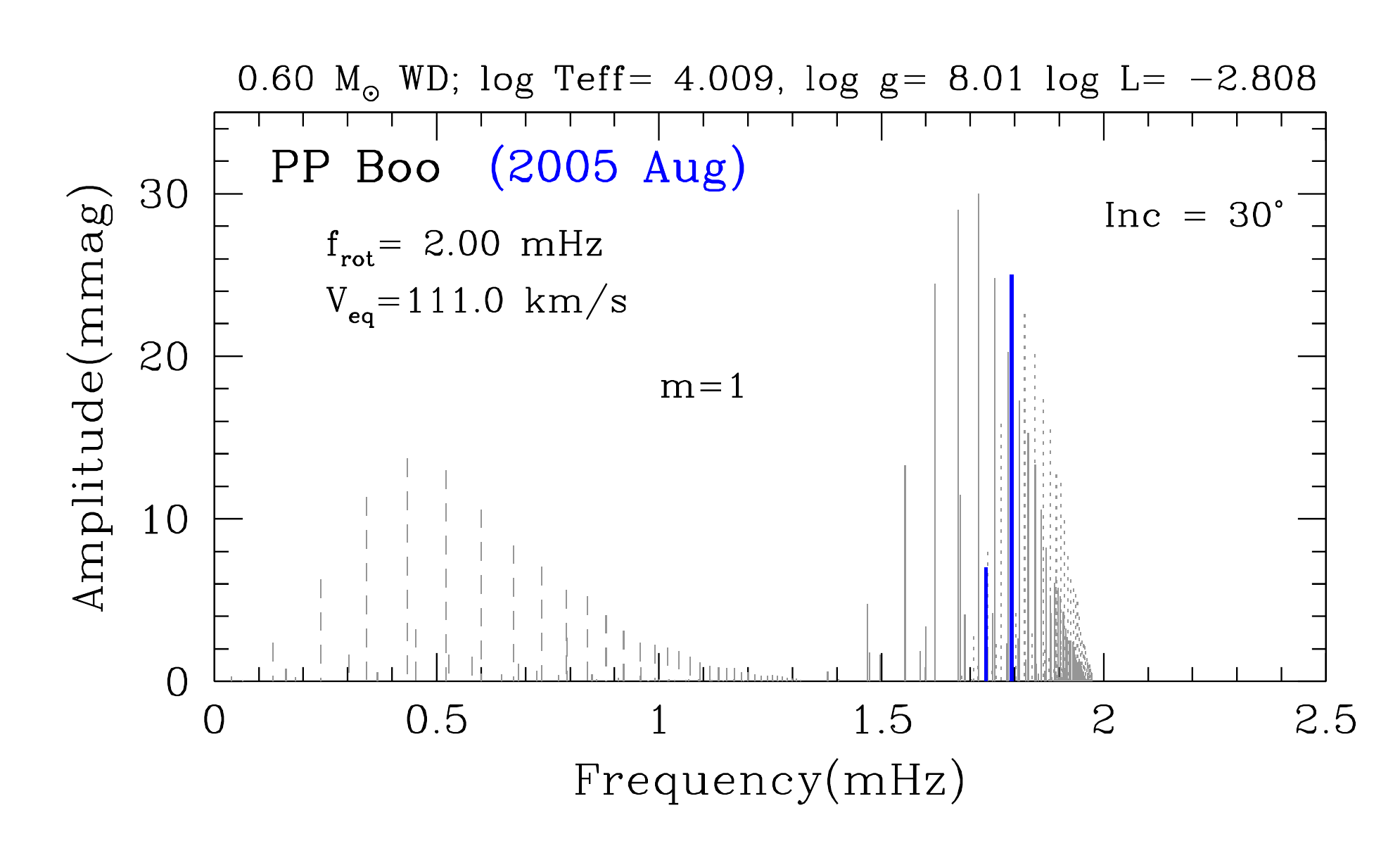}  
\caption{Pulsation frequencies/amplitudes of PP~Boo obtained by \citet{nil06} are compared with r-mode visibilities (gray lines) for a WD model with a rotation frequency of $2.00$~mHz.}
\label{fig:ppboo}
\end{figure}
PP~Boo was identified as a CV from the SDSS spectrum by \citet{szk05}.
\citet{nil06} discovered, in 2005 August, short period variations.
Fig.~\ref{fig:ppboo} shows a comparison of these frequencies with r-mode visibilities of a WD model for a rotation frequency of $2.0$~mHz.

\citet{szk10}, however, did not detect these short-period variations in their 2008 May observations, though they obtained an orbital light variation at a period of $88.8$~min from UV observations with HST. 
\citet{szk10} obtained $T_{\rm eff}=10,000\pm1000$~K for the primary WD.

\subsection{SDSS J0755 (= SDSS J075507.70+143547.6)}
\begin{figure}
\includegraphics[width=\columnwidth]{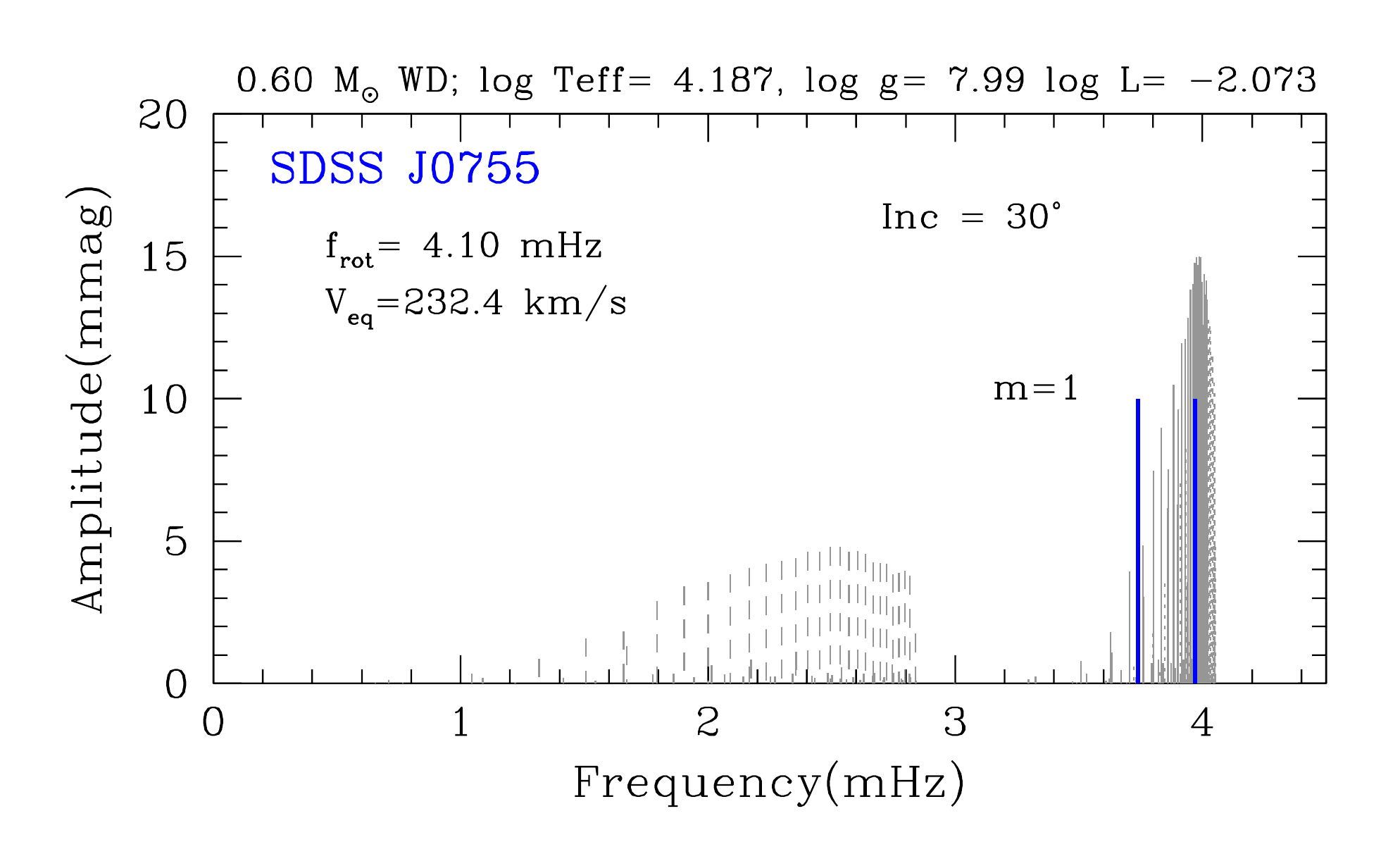}  
\caption{Frequencies of SDSS J0755 obtained by \citet{muk17} (amplitudes are assumed as $10$~mmag) are compared with r-mode ($m=1$) visibilities (gray lines) for a $0.6~M_\odot$ WD model with a rotation frequency of 4.10~mHz.}
\label{fig:j0755}
\end{figure}
SDSS~J0755 was identified as a CV from the SDSS spectrum by \citet{szk07sdss}.
\citet{muk17} discovered two periods of $265.5$ and $252$~s. 
These periodicities are compared with the visibility distribution of $m=1$ r modes in a $0.6~M_\odot$ WD model with a rotation frequency of $4.1$~mHz (244~s).
The adopted model has an effective temperature within the range $15,862 \pm 716$~K obtained by \citet{pal17}. The orbital frequency of SDSS~J0755 is $84.76$~min \citep{gae09}. 

\subsection{DY CMi (= VSX J074727.6+065050)}
\begin{figure}
\includegraphics[width=\columnwidth]{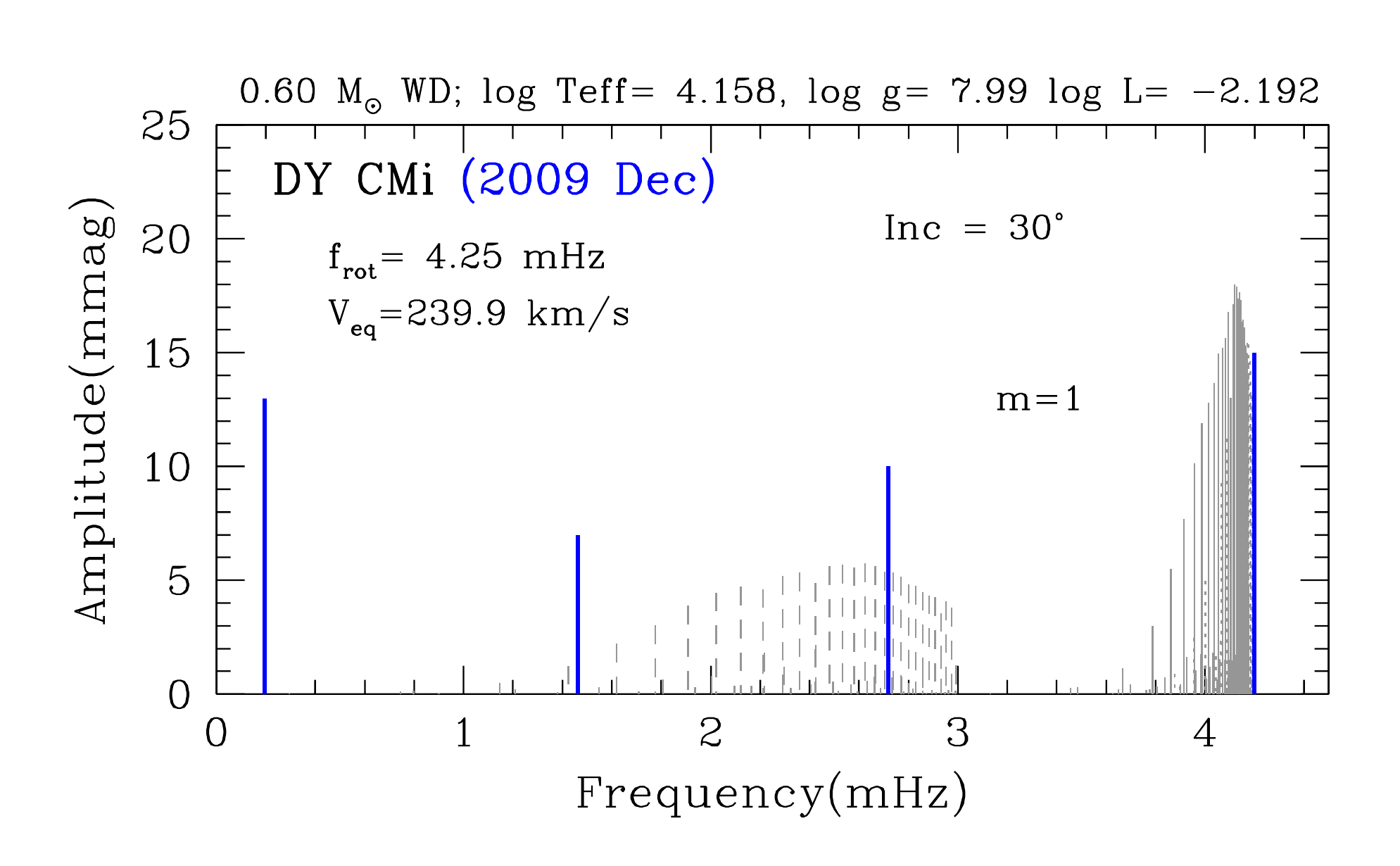}   
\caption{Frequencies/amplitudes of DY~CMi obtained in 2009 December by \citet{wou11} are compared with r-mode ($m=1$) visibilities for a $0.6~M_\odot$ WD model with a rotation frequency of $4.25$~mHz. 
The peak at $0.2$~mHz corresponds to the orbital frequency.}
\label{fig:dycmi}
\end{figure} 
The cataclysmic variable DY~CMi was discovered at an outburst in 2008 January by K. Itagaki. \citet{wou11} discovered short-period variations in 2009 December.
They also found an orbital period of $85.6$~min, and noted that the low amplitude ($\sim\!\!15$~mmag) orbital signature indicates a very low inclination \citep{wou11}.
Fig.~\ref{fig:dycmi} compares these frequencies with predicted r-mode ($m=1$) visibilities for a rotation frequency of $4.25$~mHz. Because no spectroscopic parameters of the primary white dwarf are not available, typical parameters of pulsating accreting white dwarfs \citep{szk10} are adopted as indicated in this figure.

\subsection{OV~Boo (= SDSS J150722.30+523039.8)} \label{sub:ovboo}
\begin{figure}
\includegraphics[width=\columnwidth]{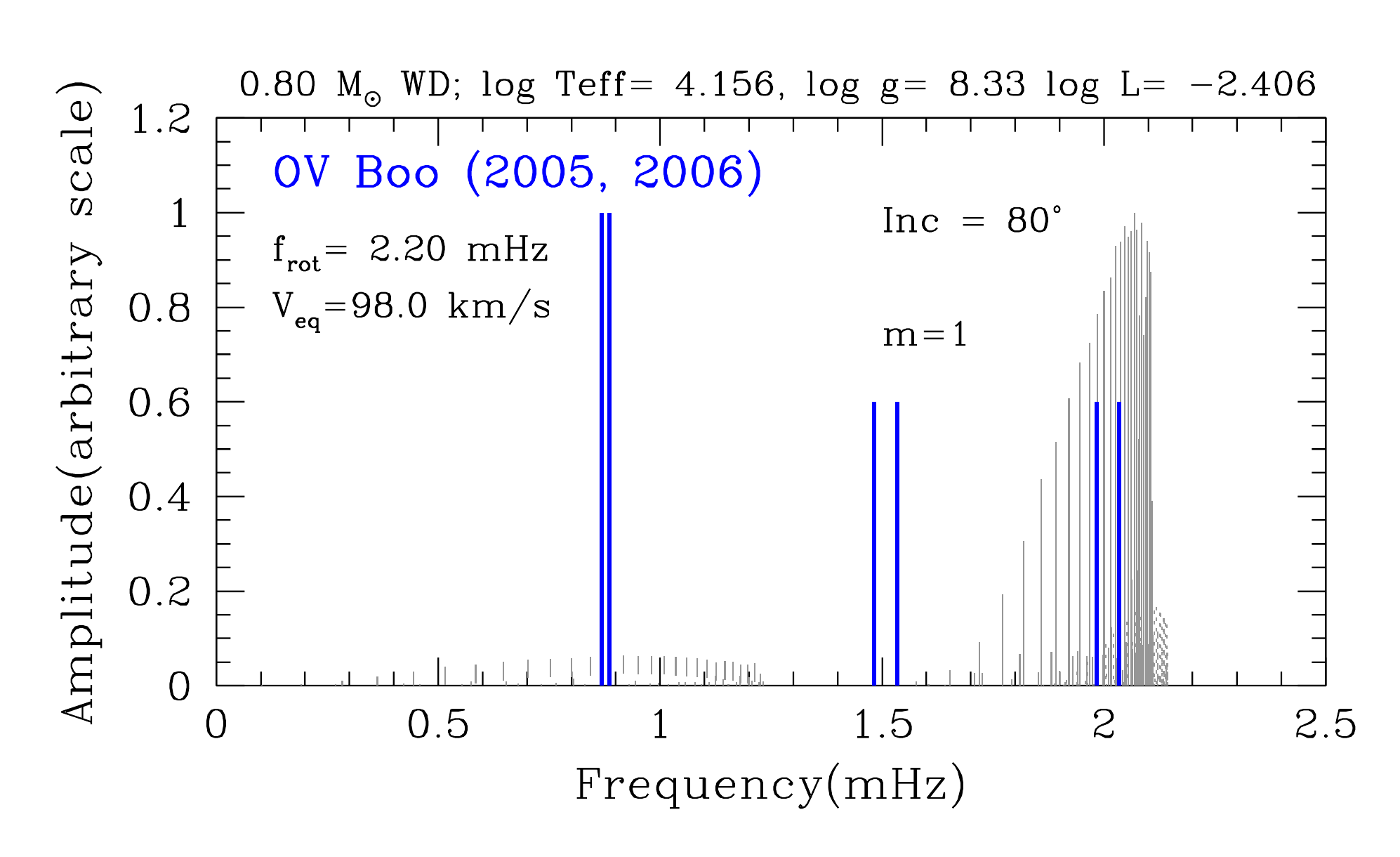}   
\caption{Frequencies of rapid variations of OV~Boo (blue solid lines) obtained by \citet{pat08} are compared with r-mode visibilities (gray lines) for a $0.8~M_\odot$ WD model with a rotation frequency of 2.2~mHz.
}  
\label{fig:ovboo}
\end{figure}
\citet{szk05} identifies OV~Boo as a CV from the SDSS spectrum and found it to be an eclipsing binary by photometry. 
The orbital period is 1.11~hr (0.25~mHz in frequency) \citep{szk05,lit07,pat08} and inclination angle is $83^\circ$ \citep{lit07,pat08,uth11}.
The orbital period is shortest among known CV binaries. 
\citet{uth11} analysed HST UV spectra for OV~Boo to obtain $T_{\rm eff} = 14200 \pm 500$~K, $\log g=8.2\pm0.3$, and a $V\sin i = 180\pm20$~km~s$^{-1}$.

\citet{pat08} found short-period pulsations of 77--75~d$^{-1}$ (0.89--0.87~mHz), 133--128d$^{-1}$ (1.53--1.48~mHz), and 176--171~d$^{-1}$ (2.04--1.98~mHz) in 2005 and 2006 observations.
\citet{pat08} noted the frequency difference between the higher two frequency groups to be equal to twice of the orbital frequency.
  
Fig.~\ref{fig:ovboo} compares these frequencies with r-mode visibility distribution for a $0.8~M_\odot$ model with a rotation frequency of 2.2~mHz.
This model fits the group at $\sim\!\!2$~mHz  to $m=1$ even ($k=-2$) r modes, assuming the group at $\sim\!\!1.5$~mHz to be combination frequencies with the orbital frequency.
If the latter group were r modes instead, the rotation frequency to fit  would be reduced to $\sim\!\!1.7$~mHz.
In this case, however, the group at $\sim\!\!0.9$~mHz would get outside of the frequency range of $m=1~(k=-1)$ r modes. 
So, the former model with faster rotation is preferable, although the predicted surface rotation velocity 98~km~s$^{-1}$ is still slower than the spectroscopic velocity obtained by \citet{uth11}.

OV~Boo underwent an outburst in 2017 March \citep{tan18}. It would be interesting to see how the pulsation frequencies have been shifted.


\bsp	
\label{lastpage}
\end{document}